\documentclass[reprint,amsmath,amssymb,aps,footinbib]{revtex4-2}  
\usepackage{dcolumn}
\usepackage{color}
\usepackage{bm}
\usepackage{xspace}
\usepackage{graphicx} 

\newcommand{\tm}[1]{\textcolor{black}{#1}}
\newcommand{\be}{\begin{equation}}
\newcommand{\ee}{\end{equation}}
\newcommand{\bea}{\begin{eqnarray}}
\newcommand{\eea}{\end{eqnarray}}
\newcommand{\beaa}{\begin{eqnarray*}}
\newcommand{\eeaa}{\end{eqnarray*}}

\begin{document}

\title{
  Angular Dependence of Specific Heat and Magnetization Effects in the Kitaev Model
}

\author{Takao Morinari}
\email{morinari.takao.5s@kyoto-u.ac.jp}
\affiliation{Course of Studies on Materials Science,
  Graduate School of Human and Environmental Studies,
  Kyoto University, Kyoto 606-8501, Japan
}

\author{Hibiki Takegami}
\email{takegami.hibiki.64h@st.kyoto-u.ac.jp}
\affiliation{Course of Studies on Materials Science,
  Graduate School of Human and Environmental Studies,
  Kyoto University, Kyoto 606-8501, Japan
}

\date{\today}

\begin{abstract}
  \tm{
    We investigate the effect of a magnetic field on the Kitaev model using the equation of motion approach for the spin Green’s function, considering both the case of suppressed magnetization ($m = 0$) and finite magnetization ($m \neq 0$). When magnetization is suppressed, the specific heat exhibits a clear $60^\circ$ periodicity in its angular dependence, with the locations of maxima and minima consistent with recent experimental observations in $\alpha$-RuCl$_3$. A qualitative difference in their temperature dependence is observed: the minima show gap-like behavior that may signal Majorana gap formation due to time-reversal symmetry breaking, while the maxima do not exhibit the expected gapless Majorana fermion signature. In addition, a linear-in-field effect---distinct from magnetization---emerges, with the characteristic temperature below which angular dependence appears increasing linearly with the magnetic field. Importantly, this directional dependence becomes quantitatively significant only at very low temperatures.
    When finite magnetization is included, the angular dependence of the specific heat remains, and the qualitative behavior is similar to the $m = 0$ case: the minima continue to exhibit gap-like features, while the maxima do not show signatures of gapless Majorana fermions. These results suggest that suppressing magnetization alone is insufficient to realize quantum spin liquid behavior in the Kitaev model under a magnetic field.
}
\end{abstract}

\maketitle


\section{Introduction}
The Kitaev model \cite{Kitaev2006}, a paradigmatic example of exactly solvable quantum spin liquid systems, has attracted significant attention in condensed matter physics due to its potential to host non-Abelian anyons and topologically protected quantum states \cite{Kitaev2003, Nayak2008, Alicea2012}. One of the most compelling approaches to studying the Kitaev model is through the formalism of Majorana fermions \cite{Kitaev2006, Feng2007, Chen2007, Chen2008a}, which maps spin degrees of freedom onto itinerant and localized fermionic ones, enabling an exact solution at zero temperature.

The physics of the Kitaev model can be explored in real materials \cite{Winter2017Kitaev, Takagi2019, Trebst2022, Kim2022}, where the dominant interaction is the Kitaev interaction mediated via the Jackeli-Khaliullin mechanism \cite{Jackeli2009}. Among these materials, $\alpha$-RuCl$_3$ \cite{Maksimov2020} has emerged as one of the most studied examples. This material exhibits a zigzag magnetic order \cite{Johnson2015, Sears2015} due to non-Kitaev interactions \cite{Winter2016, Do2017, Rau2014}. This magnetic order is suppressed by applying a magnetic field of approximately $7$T \cite{Yadav2016, Banerjee2018}, leading to the emergence of a paramagnetic state, where half-quantized thermal Hall conductivity has been observed \cite{Kasahara2018, Yokoi2021, Yamashita2020, Bruin2022}. This half-integer quantization is often attributed to the chiral edge modes of Majorana fermions, with the itinerant Majorana fermions in the B-phase becoming gapped under an applied magnetic field \cite{Kitaev2006}. The magnetic field angle dependence of heat capacity \cite{Tanaka2022} further supports this gap in the Majorana fermions. However, the presence of the chiral Majorana edge mode with a well-defined Chern number remains unsettled, as experiments \cite{Czajka2021, Czajka2022} report that the thermal Hall conductivity does not show any plateau.

\tm{
  To theoretically investigate the Kitaev spin liquid (KSL) phase in real materials, it is crucial to account for the effects of a magnetic field \cite{Janssen2019, Das2024}, as magnetic ordering must be suppressed to access the spin-liquid regime. Exact diagonalization studies on finite clusters \cite{Hickey2019, Kaib2019} have provided evidence for the KSL under magnetic fields, using indicators such as the second derivative of the ground-state energy and fidelity susceptibility to infer phase boundaries. Similarly, density matrix renormalization group studies \cite{Zhu2018, Jiang2019, Patel2019, Gohlke2018} have examined the KSL and related phases in the presence of additional interactions, for both ferromagnetic and antiferromagnetic Kitaev couplings. These finite-size approaches offer valuable insight into the nature of the phase and its response to perturbations. Nevertheless, as highlighted in tensor network studies \cite{Lee2020}, taking the thermodynamic limit is essential for a definitive identification of the KSL, particularly to capture properties such as long-range entanglement and topological degeneracy that cannot be fully resolved in small systems.
  }

  In this paper, we employ the equation of motion approach for the spin Green's function,
  a non-perturbative method free from finite-size effects, making it well-suited
  for investigating properties in the thermodynamic limit.
  We previously applied this approach to the pure Kitaev model \cite{Takegami2024, Takegami2025},
  obtaining results largely consistent with Majorana-based numerical simulations \cite{Motome2020}.
  Here, we extend our analysis to include a magnetic field, focusing on signatures of Majorana fermions
  while formulating the problem directly in terms of spin operators, rather than relying on the Majorana representation.  
  Since our primary interest is in the gap formation in the Majorana fermion spectrum,
  we analyze the magnetic field angle dependence of the specific heat, comparing our theoretical results
  with experimental observations \cite{Tanaka2022}.
  \tm{
    We find that the directional dependence of the specific heat under a magnetic field is qualitatively consistent with experimental observations.
    In the case of suppressed magnetization, the temperature dependence of the specific heat exhibits a gap-like feature when the magnetic field is applied along specific directions.
    This behavior agrees with experimental findings and shows a qualitative difference compared to directions rotated by $30^\circ$.
    However, this directional dependence becomes quantitatively significant only at very low temperatures.
    Importantly, suppressing magnetization alone is not sufficient
    to isolate the third-order term in the magnetic field, which plays a central role in the physics of the KSL.
    In fact, the characteristic temperature below which the directional difference becomes apparent increases linearly with the magnetic field strength.
    When finite magnetization is included, the angular dependence of the specific heat remains, and the behaviors of the minima and maxima
    are qualitatively similar to those in the suppressed magnetization case:
    the minima continue to exhibit gap-like features, while the maxima do not show signatures of gapless Majorana fermions.
Within our approach, the KSL is not stabilized in the ferromagnetic Kitaev model under a magnetic field.
  }

The remainder of the paper is organized as follows.
In Sec.~\ref{sec:formalism}, we introduce the equation of motion approach
for the Kitaev model under a magnetic field.
\tm{
  In Sec.~\ref{sec:m0}, we present results for the case where magnetization is suppressed,
  showing that the specific heat exhibits
  a $60^\circ$
  periodicity in its angular dependence.
  The positions of the maxima and minima are consistent with recent experimental
  observations in
  $\alpha$-RuCl$_3$.
  The temperature dependence at the minima suggests the presence of an energy gap,
  while the behavior at the maxima does not, and deviates from that expected
  for gapless Majorana fermions.
  In Sec.~\ref{sec:m_finite}, we examine the case with finite magnetization.
  Although the angular dependence of the specific heat remains qualitatively similar, the minima continue to exhibit gap-like behavior, while the maxima do not show signatures of gapless Majorana fermions, as in the suppressed magnetization case.
}
Finally, Sec.~\ref{sec:summary} summarizes our findings,
and Appendix~\ref{app:algorithm} details the algorithm for solving the self-consistent equations.

\section{Formalism}
\label{sec:formalism}
\begin{figure}[htbp]
  \includegraphics[width=0.6 \linewidth, angle=0]{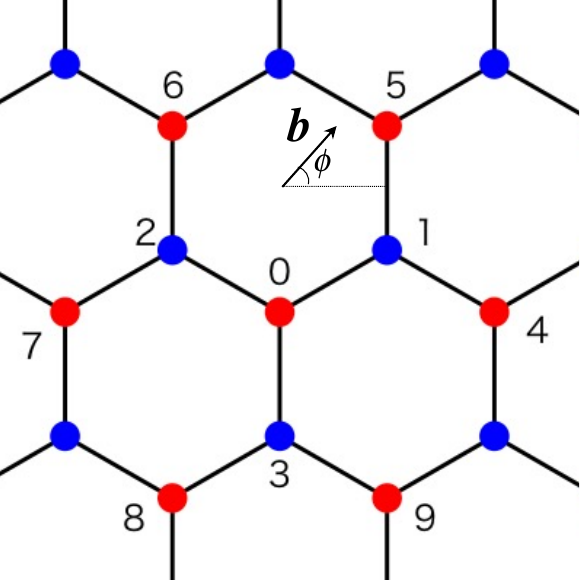}
  \caption{
    \label{fig:lattice_site_indices}
Honeycomb lattice structure in the Kitaev model.  
To aid in the presentation of the Green's function formalism and the derivation of the equations of motion, several lattice sites are numbered for clarity. The arrow indicates the direction of the applied magnetic field ${\bm b}$. The angle $\phi$ is measured from the horizontal axis, as shown in the figure.
  }
\end{figure}

We consider the Kitaev model on a honeycomb lattice, as depicted in Fig.~\ref{fig:lattice_site_indices}, under the influence of a uniform magnetic field ${\bm b} = (b_x, b_y, b_z)$. The Hamiltonian is given by
\be
H =  - \sum\limits_{\gamma  = x,y,z} {{J_\gamma }\sum\limits_{{{\left\langle {i,j} \right\rangle }_\gamma }} {S_i^\gamma S_j^\gamma } }  - \sum\limits_j
{\left( {{b_x}S_j^x + {b_y}S_j^y + {b_z}S_j^z} \right)},
\label{sec:formalism:eq:H}
\ee
where $S_j^\gamma$ represents the $\gamma$-component of the spin-$1/2$ operator at site $j$
and $\langle i,j \rangle_\gamma$ refers to the nearest-neighbor sites $i$ and $j$ connected
by a bond in the $\gamma$-direction.
For example, in Fig.~\ref{fig:lattice_site_indices}, the bond connecting site 0 and site 1 is an $x$-bond, the bond between site 0 and site 2 is a $y$-bond, and the bond between site 0 and site 3 is a $z$-bond. The other $\gamma$-bonds are parallel to these, respectively. For convenience, the lattice sites on the honeycomb lattice are labeled as shown in Fig.~\ref{fig:lattice_site_indices}.

To investigate the spin correlations in the Kitaev model, we define the following Matsubara Green's function:
\be
G_{0,j}^{{\gamma _1}{\gamma _2}}\left( \tau  \right) =  - \left\langle {{T_\tau }S_0^{{\gamma _1}}\left( \tau  \right)S_j^{{\gamma _2}}\left( 0 \right)} \right\rangle
\equiv {\left\langle {{S_0^{{\gamma _1}}}}
  \middle|
 {{S_j^{{\gamma _2}}}} \right\rangle _\tau },
\ee
where $\tau$ denotes the imaginary time, and $T_\tau$ is the imaginary-time ordering operator.
Hereafter, for any operator $A$, we define its imaginary-time dependence as
\be
A(\tau) = e^{\tau H} A e^{-\tau H}.
\ee  
We define the Fourier transform of
$G_{0,j}^{{\gamma _1}{\gamma _2}}\left( \tau  \right)$
as
\be
G_{0,j}^{{\gamma _1}{\gamma _2}}\left( {i{\omega _n}} \right) = \int_0^\beta  {d\tau } {e^{i{\omega _n}\tau }}G_{0,j}^{{\gamma _1}{\gamma _2}}\left( \tau  \right) \equiv {\left\langle {{S_0^{{\gamma _1}}}}
  \middle|
         {{S_j^{{\gamma _2}}}} \right\rangle _{i{\omega _n}}}.
\ee
Here, $\beta = 1/T$ is the inverse temperature, where $T$ denotes the temperature,
and the Boltzmann constant is set to $k_{\rm B} = 1$.
The bosonic Matsubara frequency is given by $\omega_n = 2\pi n / \beta$, where $n$ is an integer.

To analyze this Green's function, we consider its equation of motion, given by \cite{Kondo1972}:
\be
i{\omega _n}{\left\langle {{S_0^{{\gamma _1}}}}
  \middle|
      {{S_j^{{\gamma _2}}}} \right\rangle _{i{\omega _n}}}
= {\left\langle {{
      \left[ {S_0^{{\gamma _1}},H} \right] }}
    \middle|
        {{S_j^{{\gamma _2}}}} \right\rangle _{i{\omega _n}}}
+ \left\langle {\left[ {S_0^{{\gamma _1}},S_j^{{\gamma _2}}} \right]} \right\rangle.
\label{sec:formalism:eq:EOM_1st_order}
\ee
From this equation, it is clear that this approach has a hierarchical structure. To solve this equation, we need the equation of motion of the Green's function, which appears as the first term on the right-hand side. This equation of motion is obtained by replacing $S_0^{\gamma_1}$ with $[S_0^{\gamma_1}, H]$ in Eq.~(\ref{sec:formalism:eq:EOM_1st_order}). This process generates additional Green's functions.
To close the set of equations, we need to introduce an approximation. Below, we will apply the Tyablikov decoupling approximation\cite{Tyablikov1962}
to achieve this.

Computing $\left[ S_0^{\gamma_1}, H \right]$ in Eq.~(\ref{sec:formalism:eq:EOM_1st_order}), where the Hamiltonian is given by Eq.~(\ref{sec:formalism:eq:H}), we obtain, for the case of $\gamma_1 = \gamma_2 = x$:
\begin{align}
  i\omega_n \left\langle S_0^x \middle| S_j^x \right\rangle_{i{\omega _n}}
  &=  - i J_y \left\langle S_0^z S_2^y \middle| S_j^x \right\rangle_{i{\omega _n}}
  + i J_z \left\langle S_0^y S_3^z \middle| S_j^x \right\rangle_{i{\omega _n}} \nonumber \\
  & - i b_y \left\langle S_0^z \middle| S_j^x \right\rangle_{i{\omega _n}}
  + i b_z \left\langle S_0^y \middle| S_j^x \right\rangle_{i{\omega _n}}.
  \label{sec:formalism:eq:Gxx_0j}
\end{align}
This is referred to as the first-order equation of motion.
Next, we obtain the second equation of motion for the first and second terms on the right-hand side of Eq.~(\ref{sec:formalism:eq:Gxx_0j}) as follows:
\begin{align}
  & i{\omega _n}\left\langle S_0^zS_2^y\middle| S_j^x \right\rangle_{i{\omega _n}}
  \nonumber \\  
  &=  - i{J_x}\left\langle S_0^yS_1^xS_2^y\middle| S_j^x \right\rangle_{i{\omega _n}}
  - i{J_z}\left\langle S_0^zS_2^xS_6^z\middle| S_j^x \right\rangle_{i{\omega _n}}
  \nonumber \\
  & + i{J_x}\left\langle S_0^zS_2^zS_7^x\middle| S_j^x \right\rangle_{i{\omega _n}}
  + \frac{{i{J_y}}}{4}\left\langle S_0^x\middle| S_j^x \right\rangle_{i{\omega _n}}  
  \nonumber \\
  &  + i{b_x}\left\langle S_0^zS_2^z\middle| S_j^x \right\rangle_{i{\omega _n}}
  - i{b_x}\left\langle S_0^yS_2^y\middle| S_j^x \right\rangle_{i{\omega _n}}
  \nonumber \\  
  & + i{b_y}\left\langle S_0^xS_2^y\middle| S_j^x \right\rangle_{i{\omega _n}}
  - i{b_z}\left\langle S_0^zS_2^x\middle| S_j^x \right\rangle_{i{\omega _n}}
  \nonumber \\
  &  
  + i{\delta _{0,j}}\left\langle {S_0^yS_2^y} \right\rangle
  - i{\delta _{2,j}}\left\langle {S_0^zS_2^z} \right\rangle,
\end{align}  
and
\begin{align}
  & i\omega_n \left\langle  S^y_{0} S^z_{3} \middle|  S^x_{j} \right\rangle_{i{\omega _n}}
  \nonumber \\  
  &=  i J_{x}\left\langle S^z_{0}S^x_{1}S^z_{3} \middle|  S^x_{j} \right\rangle_{i{\omega _n}}
  - i J_{x}\left\langle S^y_{0}S^y_{3}S^x_{8} \middle|  S^x_{j} \right\rangle_{i{\omega _n}}
  \nonumber \\
  & + i J_{y}\left\langle S^y_{0}S^x_{3}S^y_{9} \middle|  S^x_{j} \right\rangle_{i{\omega _n}}
  - \frac{i J_{z}}{4}\left\langle S^x_{0} \middle|  S^x_{j} \right\rangle_{i{\omega _n}}
  \nonumber \\
  & + i b_{x}\left\langle S^z_{0}S^z_{3} \middle|  S^x_{j} \right\rangle_{i{\omega _n}}
  - i b_{z}\left\langle S^x_{0}S^z_{3} \middle|  S^x_{j} \right\rangle_{i{\omega _n}}
  \nonumber \\
  & - i b_{x}\left\langle S^y_{0}S^y_{3} \middle|  S^x_{j} \right\rangle_{i{\omega _n}}
  + i b_{y}\left\langle S^y_{0}S^x_{3} \middle|  S^x_{j} \right\rangle_{i{\omega _n}}
  \nonumber \\
  &
  - i\delta_{0,j}\left\langle S^z_{0}S^z_{3}\right\rangle
  + i\delta_{3,j}\left\langle S^y_{0}S^y_{3}\right\rangle.
\end{align}
By combining these two Green's functions, we define the following Green's function:
\be
F_{0,j}^{xx}\left( {i{\omega _n}} \right)
=  - i{J_y}\left\langle S_0^zS_2^y\middle| S_j^x \right\rangle_{i{\omega _n}}
+ i{J_z}\left\langle S_0^yS_3^z \middle| S_j^x \right\rangle_{i{\omega _n}},
\ee
with its corresponding equation of motion:
\begin{align}
  & i{\omega _n}F_{0,j}^{xx}\left( {i{\omega _n}} \right)
  \nonumber \\ &
  =  - {J_x}{J_y}{\left\langle S_0^yS_1^xS_2^y\middle| S_j^x \right\rangle _{i{\omega _n}}}
  - {J_y}{J_z}{\left\langle S_0^zS_2^xS_6^z\middle| S_j^x \right\rangle _{i{\omega _n}}}
  \nonumber \\ &  
  + {J_x}{J_y}{\left\langle S_0^zS_2^zS_7^x\middle| S_j^x \right\rangle _{i{\omega _n}}}
  - {J_x}{J_z}{\left\langle S_0^zS_1^xS_3^z\middle| S_j^x \right\rangle _{i{\omega _n}}}
  \nonumber \\ &    
  + {J_x}{J_z}{\left\langle S_0^yS_3^yS_8^x\middle| S_j^x \right\rangle _{i{\omega _n}}}
  - {J_y}{J_z}{\left\langle S_0^yS_3^xS_9^y\middle| S_j^x \right\rangle _{i{\omega _n}}}
  \nonumber \\ &
  - {J_y}{b_x}{\left\langle S_0^yS_2^y\middle| S_j^x \right\rangle _{i{\omega _n}}}
  + {J_y}{b_y}{\left\langle S_0^xS_2^y\middle| S_j^x \right\rangle _{i{\omega _n}}}
  \nonumber \\ &  
  - {J_z}{b_x}{\left\langle S_0^zS_3^z\middle| S_j^x \right\rangle _{i{\omega _n}}}
  + {J_z}{b_z}{\left\langle S_0^xS_3^z\middle| S_j^x \right\rangle _{i{\omega _n}}}
  \nonumber \\ &    
  + {J_y}{b_x}{\left\langle S_0^zS_2^z\middle| S_j^x \right\rangle _{i{\omega _n}}}
  - {J_y}{b_z}{\left\langle S_0^zS_2^x\middle| S_j^x \right\rangle _{i{\omega _n}}}
   \nonumber \\ &
   + {J_z}{b_x}{\left\langle S_0^yS_3^y\middle| S_j^x \right\rangle _{i{\omega _n}}}
   - {J_z}{b_y}{\left\langle S_0^yS_3^x\middle| S_j^x \right\rangle _{i{\omega _n}}}
    \nonumber \\ &   
   + \frac{{J_y^2 + J_y^2}}{4}{\left\langle S_0^x\middle| S_j^x \right\rangle _{i{\omega _n}}}
    + {J_y}{\delta _{0,j}}\left\langle S_0^yS_2^y \right\rangle
    - {J_y}{\delta _{2,j}}\left\langle S_0^zS_2^z \right\rangle
    \nonumber \\ &       
    + {J_z}{\delta _{0,j}}\left\langle S_0^zS_3^z \right\rangle
    - {J_z}{\delta _{3,j}}\left\langle S_0^yS_3^y \right\rangle.
    \label{sec:formalism:eq:Fxx_EOM}
\end{align}

We now apply the decoupling approximation to the four-spin Green's functions appearing on the right-hand side of Eq.~(\ref{sec:formalism:eq:Fxx_EOM}). There are several ways to approximate these Green's functions in terms of the product of a spin correlation function and a Green's function with fewer spin. Our strategy is to choose the decoupling method that retains the strongest correlations.
For instance, we approximate the Green's function
${\left\langle S_0^y S_1^x S_2^y \middle| S_j^x \right\rangle _{i{\omega _n}}}$  
in the following way:
\be
   {\left\langle S_0^yS_1^xS_2^y \middle| S_j^x \right\rangle _{i{\omega _n}}}
   \simeq \alpha \left\langle {S_0^yS_2^y} \right\rangle
          {\left\langle S_1^x \middle| S_j^x \right\rangle _{i{\omega _n}}}.
          \ee
          \tm{
            Here, we have introduced
            the parameter $\alpha$ \cite{Kondo1972, Shimahara1991} to
            improve the accuracy of the approximation.
            In the high-temperature limit, $\alpha \to 1$,
            indicating that the decoupling becomes exact in this regime.
            However, at lower temperatures, the decoupling is no longer exact.
            The parameter $\alpha$ is thus introduced as a correction factor
            to account for this deviation and enhance
            the validity of the approximation.
            }
          While there are alternative ways to extract spin correlation functions, the correlation function  
$\left\langle {S_0^y S_2^y} \right\rangle$  
is dominant compared to other correlation functions. 
Furthermore, the Green's function  
${\left\langle S_1^x \middle| S_j^x \right\rangle _{i{\omega _n}}}$  
describes the largest correlation in the Kitaev model when $j=0$ or $j=1$.  
For other Green's functions,
a similar decoupling approximation can be applied, though the correlations
are weaker. For instance, the Green's function  
${\left\langle S_0^zS_2^xS_6^z\middle| S_j^x \right\rangle _{i{\omega _n}}}$
can be approximated as:
\be
   {\left\langle S_0^zS_2^xS_6^z \middle| S_j^x \right\rangle _{i{\omega _n}} }
   \simeq \left\langle {S_0^zS_6^z} \right\rangle
          {\left\langle S_2^x\middle| S_j^x \right\rangle _{i{\omega _n}}}.
          \ee
Here, we note that the correlation function $\left\langle {S_0^z S_6^z} \right\rangle$ is significantly weaker compared to $\left\langle {S_0^y S_2^y} \right\rangle$. Additionally, Green's functions containing three spins appear in the formalism, but their inclusion would correspond to higher-order contributions. Therefore, we restrict our analysis to second-order approximation and neglect the higher-order contributions. Furthermore, in the self-consistent equations below, we set $j=0$ or $j=1$. Within the Kitaev interaction framework, these Green's functions describe weaker spin-spin correlations than $\left\langle S_0^x \middle|S_j^x \right\rangle_{i{\omega _n}}$, $\left\langle S_0^zS_2^y \middle|S_j^x \right\rangle_{i{\omega _n}}$, and $\left\langle S_0^yS_3^z \middle|S_j^x \right\rangle_{i{\omega _n}}$. These weaker correlations are safely neglected within the second-order approximation.
Applying these approximations, we obtain:
   \begin{align}
     i{\omega _n}F_{0,j}^{xx}(i\omega_n) &\simeq
     + \frac{{J_y^2 + J_y^2}}{4}\left\langle S_0^x\middle| S_j^x \right\rangle_{i{\omega _n}}
     \nonumber \\     
     & - \alpha {J_x}\left( {{J_y}{c_y} + {J_z}{c_z}} \right)
     \left\langle S_1^x\middle| S_j^x \right\rangle_{i{\omega _n}}
     \nonumber \\
     & + \left( {{J_y}{c_y} + {J_z}{c_z}} \right){\delta _{0,j}}.
   \end{align}
   Here, the correlation functions are defined as  
   \be
   c_\gamma = \langle S_0^\gamma S_{\delta_\gamma}^{\gamma} \rangle,
   \ee
   where $\delta_x = 1$, $\delta_y = 2$, and $\delta_z = 3$.
We also derive the equation of motion for  
$\left\langle S_0^y \middle| S_j^x \right\rangle_{i{\omega _n}}$  
and  
$\left\langle S_0^z \middle| S_j^x \right\rangle_{i{\omega _n}}$,  
which appear on the right-hand side of Eq.~(\ref{sec:formalism:eq:Gxx_0j}), while incorporating the approximations above. This leads to a set of coupled equations:
   \begin{align}
     i{\omega _n}G_{0,j}^{xx}\left( {i{\omega _n}} \right) &= F_{0,j}^{xx}\left( {i{\omega _n}} \right) - i{b_y}G_{0,j}^{zx}\left( {i{\omega _n}} \right) + i{b_z}G_{0,j}^{yx}\left( {i{\omega _n}} \right)
     \\
     i{\omega _n}G_{0,j}^{yx}\left( {i{\omega _n}} \right) &=  - i{b_z}G_{0,j}^{xx}\left( {i{\omega _n}} \right) + i{b_x}G_{0,j}^{zx}\left( {i{\omega _n}} \right) - i{\delta _{0,j}}{m_z}
     \\     
     i{\omega _n}G_{0,j}^{zx}\left( {i{\omega _n}} \right) &=  - i{b_x}G_{0,j}^{yx}\left( {i{\omega _n}} \right) + i{b_y}G_{0,j}^{xx}\left( {i{\omega _n}} \right) + i{\delta _{0,j}}{m_y}
     \\     
     i{\omega _n}F_{0,j}^{xx}\left( {i{\omega _n}} \right)
     &= 
     - {J_x}\alpha \left( {{J_y}{c_y} + {J_z}{c_z}} \right)G_{1,j}^{xx}\left( {i{\omega _n}} \right)
     \nonumber \\          
     & +\frac{{J_y^2 + J_z^2}}{4}G_{0,j}^{xx}\left( {i{\omega _n}} \right)
     + {\delta _{0,j}}\left( {{J_y}{c_y} + {J_z}{c_z}} \right).
   \end{align}
Since the fourth equation contains $G_{1,j}^{xx} (i{\omega_n})$, we require its equation of motion. Expanding it up to second order in a similar manner, we obtain:
   \begin{align}
     i{\omega _n}G_{1,j}^{xx}\left( {i{\omega _n}} \right) &= F_{1,j}^{xx}\left( {i{\omega _n}} \right) - i{b_y}G_{1,j}^{zx}\left( {i{\omega _n}} \right) + i{b_z}G_{1,j}^{yx}\left( {i{\omega _n}} \right)
     \\
     i{\omega _n}G_{1,j}^{yx}\left( {i{\omega _n}} \right) &= i{b_x}G_{1,j}^{zx}\left( {i{\omega _n}} \right) - i{b_z}G_{1,j}^{xx}\left( {i{\omega _n}} \right) - i{\delta _{1,j}}{m_z}
     \\
     i{\omega _n}G_{1,j}^{zx}\left( {i{\omega _n}} \right) &=  - i{b_x}G_{1,j}^{yx}\left( {i{\omega _n}} \right) + i{b_y}G_{1,j}^{xx}\left( {i{\omega _n}} \right) + i{\delta _{1,j}}{m_y}
     \\
     i{\omega _n}F_{1,j}^{xx}\left( {i{\omega _n}} \right)
     &= 
     - {J_x}\alpha \left( {{J_y}{c_y} + {J_z}{c_z}} \right)G_{0,j}^{xx}\left( {i{\omega _n}} \right)
     \nonumber \\
     & +\frac{{J_y^2 + J_z^2}}{4}G_{1,j}^{xx}\left( {i{\omega _n}} \right)
     + {\delta _{1,j}}\left( {{J_y}{c_y} + {J_z}{c_z}} \right),
   \end{align}  
   where:
   \be
   F_{1,j}^{xx}\left( {i{\omega _n}} \right)
   =  - i{J_y}{\left\langle S_1^zS_4^y\middle| S_j^x \right\rangle _{i{\omega _n}}}
   + i{J_z}{\left\langle S_1^yS_5^z \middle| S_j^x \right\rangle _{i{\omega _n}}}.
   \ee
To compute $c_y$ and $c_z$, we require a similar set of equations for $G_{0,j}^{yy} (i{\omega_n})$ and $G_{0,j}^{zz} (i{\omega_n})$. These can be derived from the above equations by applying the cyclic permutation:
$x \rightarrow y \rightarrow z \rightarrow x, \quad 1 \rightarrow 2 \rightarrow 3 \rightarrow 1$.

The set of equations of motion is summarized in the following equation:
\be
i{\omega _n}{{\bm{G}}^\gamma }\left( {i{\omega _n}} \right) = {M^\gamma }{{\bm{G}}^\gamma }\left( {i{\omega _n}} \right) + {{\bm{c}}^\gamma },
\label{eq:formalism:matrix_G_form}
\ee
\begin{widetext}
where 
\begin{align}
{{\bm{G}}^x}\left( {i{\omega _n}} \right) &= {\left( {\begin{array}{*{20}{c}}
{G_{0,0}^{xx}}&{G_{0,0}^{yx}}&{G_{0,0}^{zx}}&{F_{0,0}^{xx}}&{G_{1,0}^{xx}}&{G_{1,0}^{yx}}&{G_{1,0}^{zx}}&{F_{1,0}^{xx}}
  \end{array}} \right)^{T}},
\\  
{{\bm{G}}^y}\left( {i{\omega _n}} \right) &= {\left( {\begin{array}{*{20}{c}}
{G_{0,0}^{yy}}&{G_{0,0}^{zy}}&{G_{0,0}^{xy}}&{F_{0,0}^{yy}}&{G_{2,0}^{yy}}&{G_{2,0}^{zy}}&{G_{2,0}^{xy}}&{F_{2,0}^{yy}}
\end{array}} \right)^{T}},
\\  
{{\bm{G}}^z}\left( {i{\omega _n}} \right) &= {\left( {\begin{array}{*{20}{c}}
{G_{0,0}^{zz}}&{G_{0,0}^{xz}}&{G_{0,0}^{yz}}&{F_{0,0}^{zz}}&{G_{3,0}^{zz}}&{G_{3,0}^{xz}}&{G_{3,0}^{yz}}&{F_{3,0}^{zz}}
\end{array}} \right)^{T}}.
\end{align}
Note that the dependence on $i\omega_n$ in the right-hand side of these equations is implicit.
The matrices $M^x$, $M^y$, and $M^z$ are given by
\be
   {M^x} = \left( {\begin{array}{*{20}{c}}
0&{i{b_z}}&{ - i{b_y}}&1&0&0&0&0\\
{ - i{b_z}}&0&{i{b_x}}&0&0&0&0&0\\
{i{b_y}}&{ - i{b_x}}&0&0&0&0&0&0\\
{\frac{1}{4}\left( {J_y^2 + J_z^2} \right)}&0&0&0&{ - {J_x}\alpha \left( {{J_y}{c_y} + {J_z}{c_z}} \right)}&0&0&0\\
0&0&0&0&0&{i{b_z}}&{ - i{b_y}}&1\\
0&0&0&0&{ - i{b_z}}&0&{i{b_x}}&0\\
0&0&0&0&{i{b_y}}&{ - i{b_x}}&0&0\\
{ - {J_x}\alpha \left( {{J_y}{c_y} + {J_z}{c_z}} \right)}&0&0&0&{\frac{1}{4}\left( {J_y^2 + J_z^2} \right)}&0&0&0
\end{array}} \right),
\ee
\be
   {M^y} = \left( {\begin{array}{*{20}{c}}
0&{i{b_x}}&{ - i{b_z}}&1&0&0&0&0\\
{ - i{b_x}}&0&{i{b_y}}&0&0&0&0&0\\
{i{b_z}}&{ - i{b_y}}&0&0&0&0&0&0\\
{\frac{1}{4}\left( {J_z^2 + J_x^2} \right)}&0&0&0&{ - {J_y}\alpha \left( {{J_z}{c_z} + {J_x}{c_x}} \right)}&0&0&0\\
0&0&0&0&0&{i{b_x}}&{ - i{b_z}}&1\\
0&0&0&0&{ - i{b_x}}&0&{i{b_y}}&0\\
0&0&0&0&{i{b_z}}&{ - i{b_y}}&0&0\\
{ - {J_y}\alpha \left( {{J_z}{c_z} + {J_x}{c_x}} \right)}&0&0&0&{\frac{1}{4}\left( {J_z^2 + J_x^2} \right)}&0&0&0
\end{array}} \right),
   \ee
   \be
      {M^z} = \left( {\begin{array}{*{20}{c}}
0&{i{b_y}}&{ - i{b_x}}&1&0&0&0&0\\
{ - i{b_y}}&0&{i{b_z}}&0&0&0&0&0\\
{i{b_x}}&{ - i{b_z}}&0&0&0&0&0&0\\
{\frac{1}{4}\left( {J_x^2 + J_y^2} \right)}&0&0&0&{ - {J_z}\alpha \left( {{J_x}{c_x} + {J_y}{c_y}} \right)}&0&0&0\\
0&0&0&0&0&{i{b_y}}&{ - i{b_x}}&1\\
0&0&0&0&{ - i{b_y}}&0&{i{b_z}}&0\\
0&0&0&0&{i{b_x}}&{ - i{b_z}}&0&0\\
{ - {J_z}\alpha \left( {{J_x}{c_x} + {J_y}{c_y}} \right)}&0&0&0&{\frac{1}{4}\left( {J_x^2 + J_y^2} \right)}&0&0&0
\end{array}} \right).
      \ee
The vectors ${\bm{c}}^x$, ${\bm{c}}^y$, and ${\bm{c}}^z$ are defined as      
\begin{align}
{{\bm{c}}^x} &= {\left( {\begin{array}{*{20}{c}}
0&{ - i{m_z}}&{i{m_y}}&{{J_y}{c_y} + {J_z}{c_z}}&0&0&0&0
  \end{array}} \right)^{T}},
\label{eq:cx}
\\        
{{\bm{c}}^y} &= {\left( {\begin{array}{*{20}{c}}
0&{ + i{m_z}}&{ - i{m_x}}&{{J_z}{c_z} + {J_x}{c_x}}&0&0&0&0
  \end{array}} \right)^{T}},
\label{eq:cy}
\\
{{\bm{c}}^z} &= {\left( {\begin{array}{*{20}{c}}
0&{ - i{m_y}}&{i{m_x}}&{{J_x}{c_x} + {J_y}{c_y}}&0&0&0&0
  \end{array}} \right)^{T}},
\label{eq:cz}
\end{align}  
\end{widetext}
where
\be
{m_\gamma } = \left\langle {S_0^\gamma } \right\rangle.
\ee
We note that in the paramagnetic phase, we may set
\be
\left( {{m_x},{m_y},{m_z}} \right) = \frac{{\bm{b}}}{{\left| {\bm{b}} \right|}}m.
\ee
Therefore, the quantity to be self-consistently determined is $m$,
rather than each individual component $m_\gamma$.

The self-consistent equation is conveniently expressed in terms of the imaginary-time Green's function,
\be
   {{\bm{G}}^\gamma }\left( \tau  \right) = \frac{1}{\beta }\sum\limits_{i{\omega _n}} {{e^{ - i{\omega _n}\tau }}} {{\bm{G}}^\gamma }\left( {i{\omega _n}} \right),
   \ee
   as
   \begin{align}
     {\left[ {{{\bm{G}}^\gamma }\left( {\tau  =  + 0} \right)} \right]_0}
     &= - \left\langle {S_0^\gamma S_0^\gamma } \right\rangle
     = - \frac{1}{4},
     \label{sec:formalism:eq:sc_eq0}
\\
  {\left[ {{{\bm{G}}^\gamma }\left( {\tau  =  + 0} \right)} \right]_4}
  &=  - \left\langle {S_{{\delta _\gamma }}^\gamma S_0^\gamma } \right\rangle 
  = - {c_\gamma }.
     \label{sec:formalism:eq:sc_eq_gamma}  
   \end{align}
     Note that there are six self-consistent equations,
     while the number of parameters to be determined is five: $c_x$, $c_y$, $c_z$, $\alpha$, and $m$.
     Since there is one additional equation beyond the number of unknowns,
     we must carefully select a subset of equations to match the number of parameters.
     This selection must be performed without breaking the symmetry of the system.
To facilitate this reduction, we define the following quantities:
   \begin{align}
     I_\gamma^0 \equiv{\left[ {{{\bm{G}}^\gamma }\left( {\tau  =  + 0} \right)} \right]_0} + \frac{1}{4},
     \\
     I_\gamma \equiv
     {\left[ {{{\bm{G}}^\gamma }\left( {\tau  =  + 0} \right)} \right]_4} + {c_\gamma }.
   \end{align}
For the case of $m \neq 0$, we solve the following set of five self-consistent equations:   
\begin{align}
& I_x = 0,
\\  
& I_y = 0,
\\  
& I_z = 0,
\\  
& I_x^0 + I_y^0 + I_z^0 = 0,
\\  
& I_x^0 I_y^0 + I_y^0 I_z^0 + I_z^0 I_x^0 = 0.
\end{align}
For the case of $m = 0$, we solve the first four equations from the above set.
In Appendix~\ref{app:algorithm}, we provide details on how to solve the Green's functionn equations
in the matrix form given by Eq.~(\ref{eq:formalism:matrix_G_form}).

\section{Magnetic Field Effects under Conserved $\mathbb{Z}_2$ Gauge Fluxes}
\label{sec:m0}
  \tm{
    Kitaev explored the effect of time-reversal symmetry breaking in the B-phase
    to explain the gap opening in massless Majorana fermions \cite{Kitaev2006}.
    In his approach, the magnetic field is treated perturbatively,
    and the Hilbert space is restricted to configurations
    without free $\mathbb{Z}_2$ gauge fluxes.
    This restriction causes the first-order term in the magnetic field expansion
    to vanish, with the time-reversal symmetry-breaking term emerging only at third order.
    To study this effect theoretically, it is necessary to suppress the magnetization,
    which arises at first order in the magnetic field.
    Accordingly, in this section,
    we focus on the case of suppressed magnetization by setting $m = 0$.
  }
  
  \tm{
    In our study, we focus on the properties of the spin-liquid phase, with particular emphasis on the dependence of spin correlations on the direction of the magnetic field. Therefore, in this and the following sections, we primarily consider magnetic fields smaller than $0.05J$, in order to remain within the non-spin-polarized regime \cite{Nasu2018}.
    }
  Inspired by the experimental work in \cite{Tanaka2022}, we analyze the angular dependence of the specific heat to explore how the magnetic field influences the gap formation. For $\alpha$-RuCl$_3$, the unit vectors along the a-axis and b-axis are given by
\begin{align}
{{\bm{e}}_a} &= \frac{1}{{\sqrt 6 }}\left( {1,1, - 2} \right),
\\
{{\bm{e}}_b} &= \frac{1}{{\sqrt 2 }}\left( { - 1,1,0} \right),
\end{align}
respectively. In Fig.~\ref{fig:lattice_site_indices}, the horizontal direction is parallel to ${{\bm{e}}_a}$, while the vertical direction is parallel to ${{\bm{e}}_b}$.
If we define the angle $\phi$ of the magnetic field as shown in Fig.~\ref{fig:lattice_site_indices}, then
\be
   {\bm{b}} = b\left( {{{\bm{e}}_a}\cos \phi  + {{\bm{e}}_b}\sin \phi } \right).
\ee   
Thus,
\begin{align}
{b_x} &= \left( {\frac{1}{{\sqrt 6 }}\cos \phi  - \frac{1}{{\sqrt 2 }}\sin \phi } \right)b,
\\  
{b_y} &= \left( {\frac{1}{{\sqrt 6 }}\cos \phi  + \frac{1}{{\sqrt 2 }}\sin \phi } \right)b,
\\  
{b_z} &=  - \frac{2}{{\sqrt 6 }}b\cos \phi.
\end{align}  

In solving the self-consistent equation, we first solve it at high temperature, $T \sim 10 J_{\gamma}$, where the initial values are taken from the high-temperature expansion (HTE) as follows:
\begin{align}
  {\langle S_0^xS_1^x\rangle _{{\rm{HTE}}}}
  &= \frac{{{J_x}\beta }}{{16}} + \frac{{b_x^2{\beta ^2}}}{{16}} + O(\beta^3),
\\  
  {\langle S_0^yS_2^y\rangle _{{\rm{HTE}}}} &= \frac{{{J_y}\beta }}{{16}}
  + \frac{{b_y^2{\beta ^2}}}{{16}} + O(\beta^3),
\\
  {\langle S_0^zS_3^z\rangle _{{\rm{HTE}}}} &= \frac{{{J_z}\beta }}{{16}}
  + \frac{{b_z^2{\beta ^2}}}{{16}} + O(\beta^3),
\end{align}
where $\langle ... \rangle_{\rm{HTE}}$ denotes the expectation value obtained from HTE.
  The HTE value of the parameter $\alpha$ can be determined from the HTE values of the correlation functions.
  However, since $\alpha \to 1$ as $T \to \infty$, we may simply set $\alpha_{\rm HTE} = 1$.
  \tm{
    In the following, we focus on the isotropic Kitaev interaction
    by setting $J_x = J_y = J_z = J$.
  }

\begin{figure}[htbp]
  \includegraphics[width=1 \linewidth, angle=0]{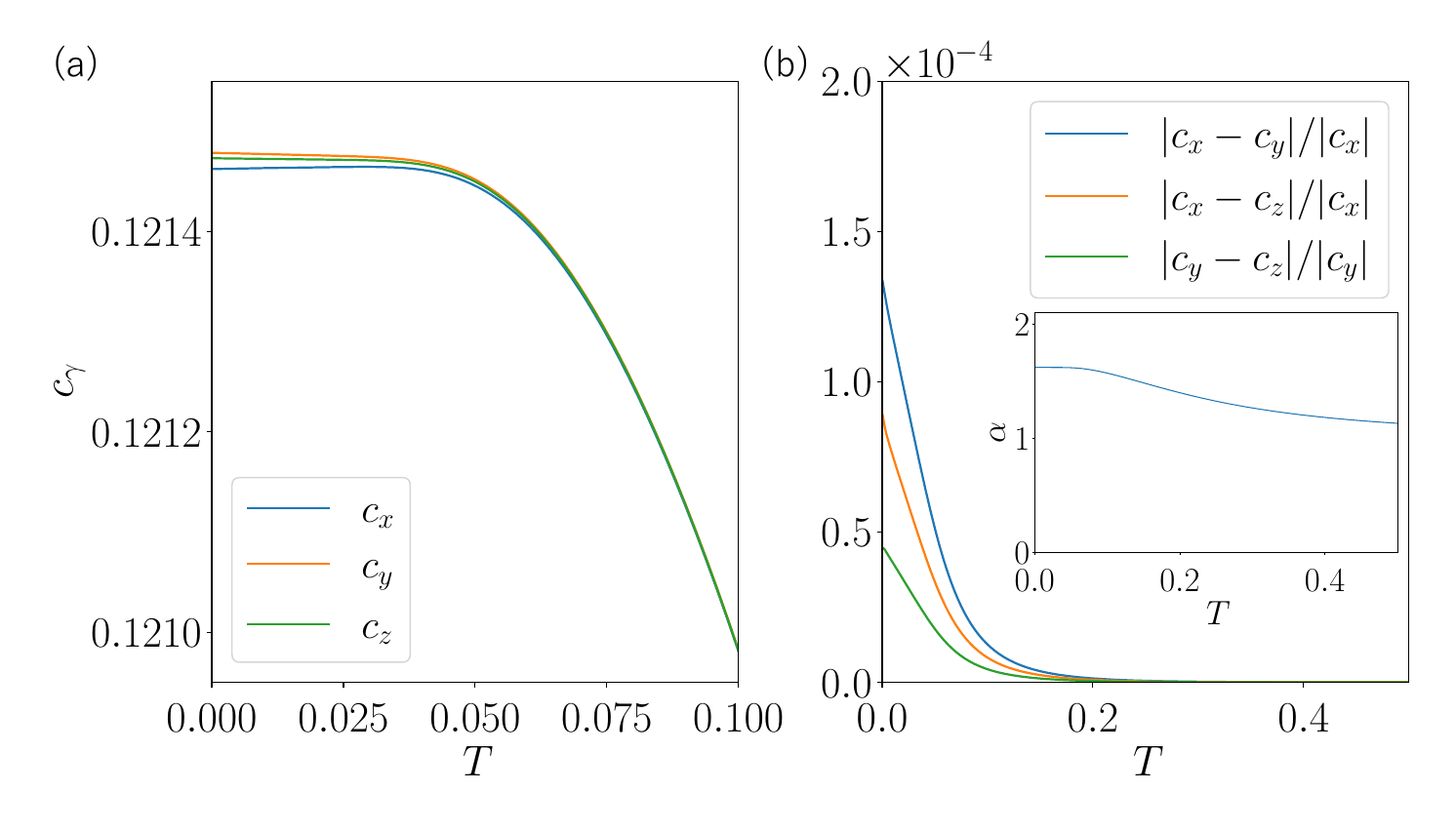}
  \caption{
    \label{fig:m0:T_c_alpha}
    (a) Temperature dependence of the spin correlation functions $c_x$, $c_y$, and $c_z$. 
    (b) Temperature dependence of the differences in the spin correlation functions.
    The inset shows the temperature dependence of the decoupling parameter $\alpha$,
    which is introduced in the decoupling approximation.  
    Both (a) and (b) are computed for
    \tm{$b = 0.01$}
    and $\phi = 40^\circ$,
    where the magnetic field components are given by
    $b_x = -1.42 \times 10^{-3}$,
    $b_y = 7.67 \times 10^{-3}$, and
    $b_z = -6.25 \times 10^{-3}$.
  }
\end{figure}

Figure~\ref{fig:m0:T_c_alpha} shows the temperature dependence of the spin correlation functions and the parameter $\alpha$ at $b=0.2$ and $\phi = 40^\circ$.  
From Fig.~\ref{fig:m0:T_c_alpha}(a), we observe that anisotropy in the correlation functions emerges at low temperatures. At temperatures above $\sim 0.2$, the correlation functions remain nearly isotropic, as confirmed by the temperature dependence of their differences shown in Fig.~\ref{fig:m0:T_c_alpha}(b).
Meanwhile, we find that $\alpha \sim 1$ at high temperatures, as shown in the inset of Fig.~\ref{fig:m0:T_c_alpha}(a). This indicates that the decoupling approximation is accurate at high temperatures.

\begin{figure}[htbp]
  \includegraphics[width=0.9 \linewidth, angle=0]{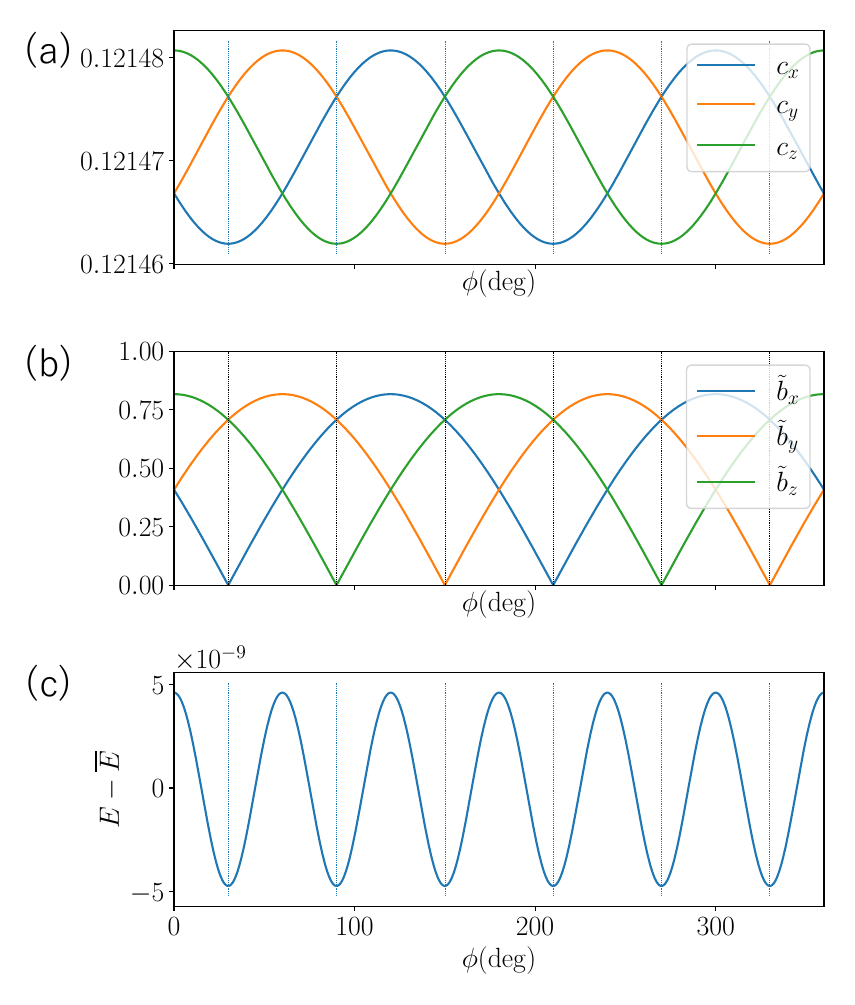}
  \caption{
    \label{fig:m0:c_E_phi_dep}
    \tm{
      (a) Angular dependence of the spin correlation functions at $b = 0.01$ and $T = 0.001$.
      (b) Angular dependence of $\tilde{b}_\gamma = |b_\gamma| / |\bm{b}|$ for $\gamma = x, y, z$. While the overall behavior of $\tilde{b}_\gamma$ is qualitatively similar to that of the spin correlation functions in (a), if the correlation functions exactly followed the angular dependence of $\tilde{b}_\gamma$, the energy $E$ would exhibit no oscillation.
      (c) Energy oscillation resulting from the deviation between the angular dependence in (a) and (b). Here, $\overline{E}$ denotes the angle-averaged energy. The vertical lines indicate the angles $\phi = 30^\circ + 60^\circ \times n$, where $n$ is an integer.
    }
  }
\end{figure}

\begin{figure}[htbp]
  \includegraphics[width=0.9 \linewidth, angle=0]{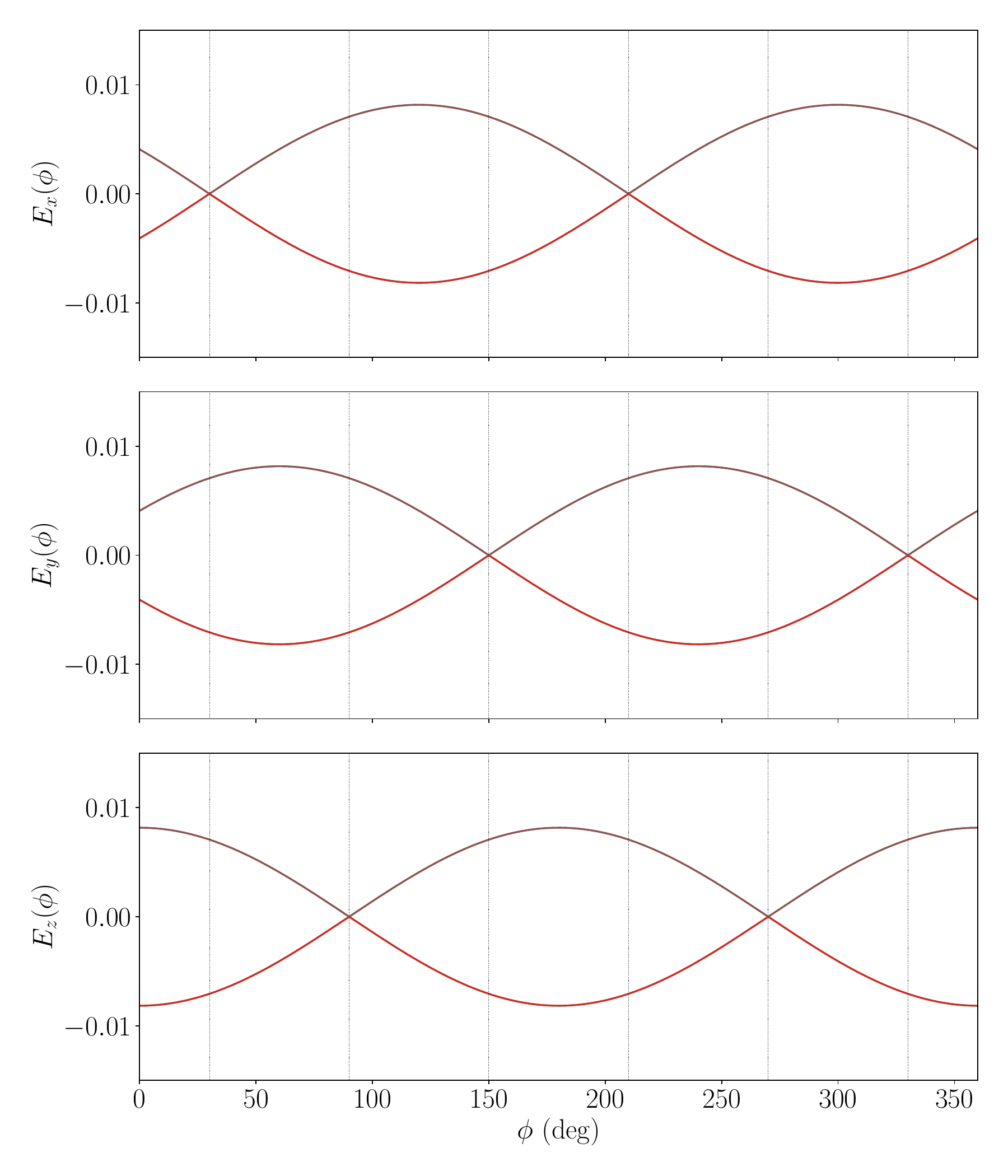}
  \caption{
    \label{fig:m0:G_poles_phi}
    \tm{
      Angular dependence of the poles of the Green's function ${\bm{G}}^\gamma(i\omega_n)$,
      denoted as $E_\gamma(\phi)$, at $b = 0.01$ and $T = 0.001$.
      There are eight poles in total.
      The four poles located near zero energy are shown.
      Zero-energy poles emerge when $b_\gamma = 0$, indicating points
      where the energy gap closes.
      Each oscillating curve near zero energy actually consists of
      two nearly degenerate lines with only a minuscule difference between them.
    }
  }
\end{figure}

\begin{figure}[htbp]
  \includegraphics[width=0.9 \linewidth, angle=0]{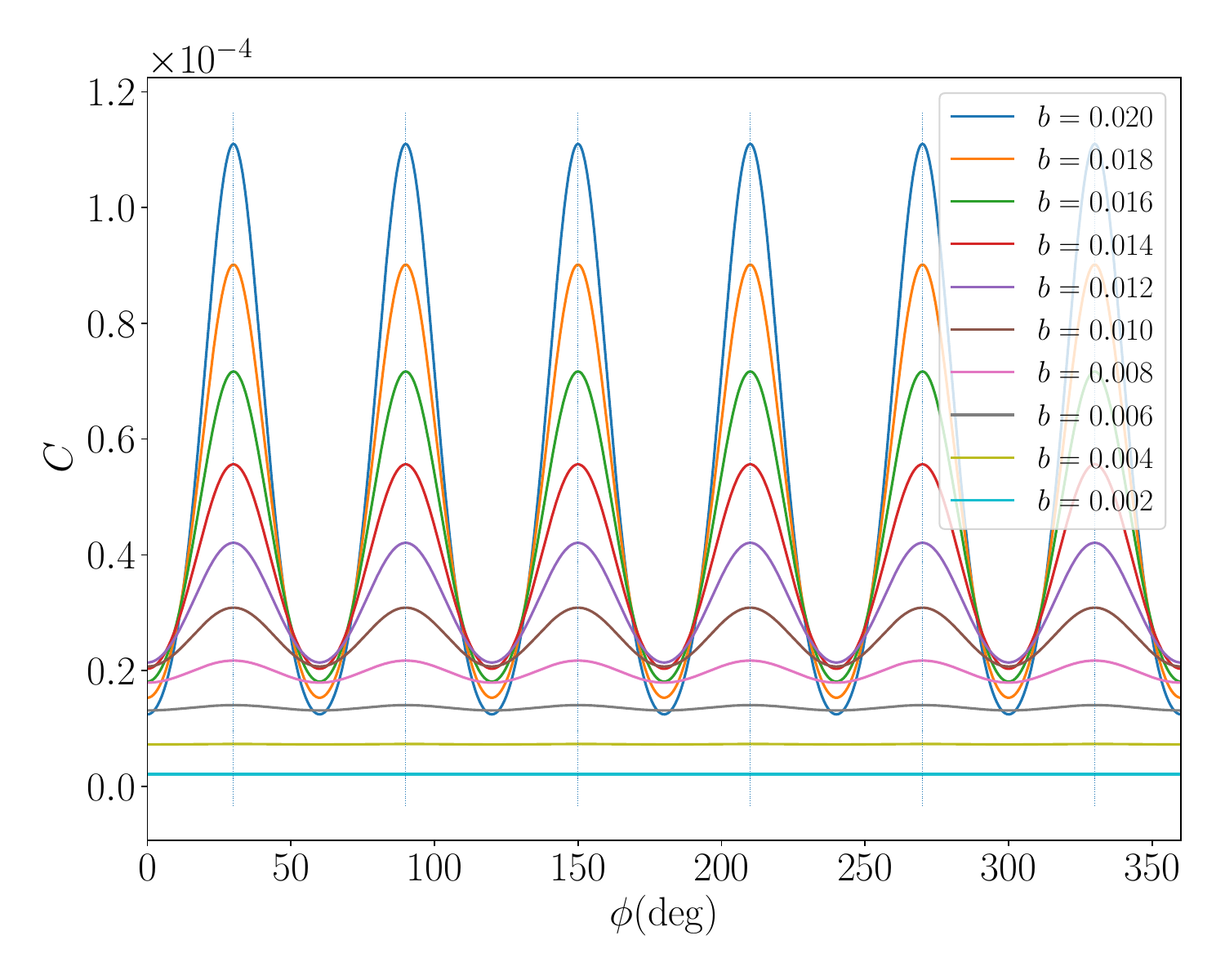}
  \caption{
    \label{fig:m0:C_phi_b}
    Angular dependence of the specific heat at
    \tm{$T = 0.001$}
    for different magnetic fields.
    The vertical lines indicate specific angles as described earlier.
  }
\end{figure}

We now investigate the angle $\phi$ dependence of the spin correlations. We start from the solution at a given temperature and $\phi$, using this solution as the initial condition, and then solve the self-consistent equations while gradually increasing $\phi$. 
Figure~\ref{fig:m0:c_E_phi_dep} shows the $\phi$ dependence of the spin correlation functions and energy at $b = 0.01$ and $T = 0.001$.
The $\phi$ dependence of the spin correlation functions, shown in Fig.~\ref{fig:m0:c_E_phi_dep}(a), can be approximately understood in terms of the quantity  
${{\tilde b}_\gamma} \equiv \left| {b_\gamma} \right|/\left| {\mathbf{b}} \right|$, which is plotted in Fig.~\ref{fig:m0:c_E_phi_dep}(b). The correlation function $c_\gamma$ reaches its minimum values when ${\tilde b}_\gamma = 0$.
The energy per site can be computed as:
\be
E =  - \frac{1}{2} J \left( c_x + c_y + c_z \right).
\ee
Figure~\ref{fig:m0:c_E_phi_dep}(c) shows the $\phi$ dependence of $E$.
We observe a clear $60^\circ$ periodicity, although the amplitude is quite small.
\tm{
  We emphasize that if the correlation functions exactly followed
  the angular dependence of $\tilde{b}_\gamma$,
  the energy $E$ would exhibit no oscillation.
  }
Figure~\ref{fig:m0:G_poles_phi} presents the $\phi$ dependence of the poles of the Green's function, ${\bm{G}}^\gamma(i\omega_n)$, denoted as $E_\gamma(\phi)$.
Notably, zero-energy poles appear in ${\bm{G}}^\gamma(i\omega_n)$ when $b_\gamma = 0$. 
The specific heat is obtained by taking the temperature derivative of the energy.
Figure~\ref{fig:m0:C_phi_b} presents the $\phi$ dependence of the specific heat
at $T = 0.001$ for different magnetic fields.
The specific heat exhibits a $60^\circ$ periodicity, and the amplitude increases with increasing magnetic field.
  This $60^\circ$ periodicity, along with the positions of the maxima and minima,
  is consistent with the experimental results reported for $\alpha$-RuCl$_3$
  in Ref.~\cite{Tanaka2022}.

\begin{figure}[htbp]
  \includegraphics[width=0.9 \linewidth, angle=0]{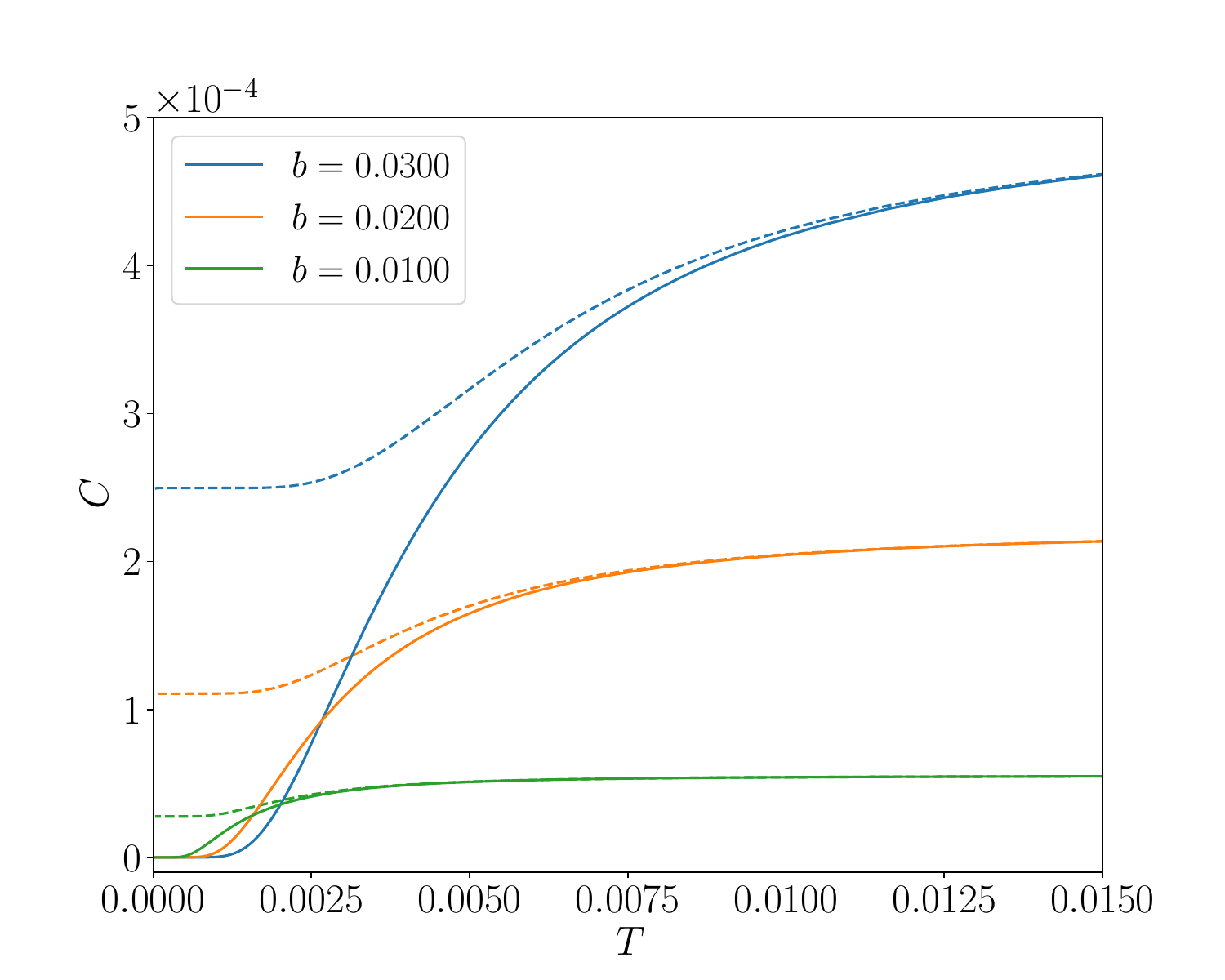}
  \caption{
    \label{fig:m0:C_over_T_b_dep}
    \tm{
      Temperature dependence of the specific heat $C$ for various magnetic field strengths
      applied along the $\phi = 0^\circ$ direction (solid lines)
    and the $\phi = 90.01^\circ$ direction (dashed lines).
    }
  }
\end{figure}

The angular dependence of the energy and specific heat implies a nontrivial interplay between the spin correlations in the Kitaev model and the magnetic field.  
From the perspective of fractionalized Majorana fermions, the gapless Majorana fermions acquire a gap in the presence of a magnetic field, and this gap is proportional to\cite{Kitaev2006}  
$\left| {{b_x}{b_y}{b_z}} \right| \propto \left| {\cos 3\phi } \right|$.
As a result, the Majorana fermions remain gapless when
$\phi = 30^\circ + 60^\circ \times n$,
while the gap is maximized at
$\phi = 60^\circ \times n$,
where $n$ is an integer.
  When the Majorana fermions are gapped, the specific heat decreases exponentially with decreasing temperature.
  In contrast, when the Majorana fermions remain gapless,
  the specific heat follows a $T^2$ dependence at low temperatures.
  \tm{
    Figure~\ref{fig:m0:C_over_T_b_dep} shows the temperature dependence
    of the specific heat $C$ for different magnetic field strengths
    applied along the $\phi = 0^\circ$ direction and the $\phi = 90.01^\circ$ direction.
    We note that spin inversion symmetry exists for each spin component
    at specific magnetic field directions.
    For example, the $z$-component exhibits this symmetry
    at angles $\phi = 90^\circ + 180^\circ \times n \equiv \phi_z(n)$,
    where $n$ is an integer.
    The $x$-component exhibits this symmetry at $\phi = \phi_z(n) - 60^\circ$,
    and the $y$-component at $\phi = \phi_z(n) - 30^\circ$.
    Because such exact symmetries are generally not preserved
    in real experimental conditions,
    we intentionally avoid these special angles in our numerical calculations.
    A qualitative difference is observed between the two cases:
    the specific heat exhibits a gap-like feature
    when the magnetic field is applied along the $\phi = 0^\circ$ direction,
    while no such behavior is seen for $\phi = 90.01^\circ$.
    A gap in the Majorana fermion spectrum is expected for $\phi = 0^\circ$,
    which is consistent with our results.
    In contrast, we do not observe any features indicative of massless Majorana fermions at $\phi = 90.01^\circ$.
    Notably, the specific heat approaches a finite value
    as $T \rightarrow 0$, which is unphysical.
    This behavior is likely due, in part, to the artificial suppression
    of magnetization, and also reflects the limited accuracy
    of our approach at low temperatures.
    This issue is discussed in more detail later
    in the context of finite magnetization.
    }

\begin{figure}[htbp]
  \includegraphics[width=0.9 \linewidth, angle=0]{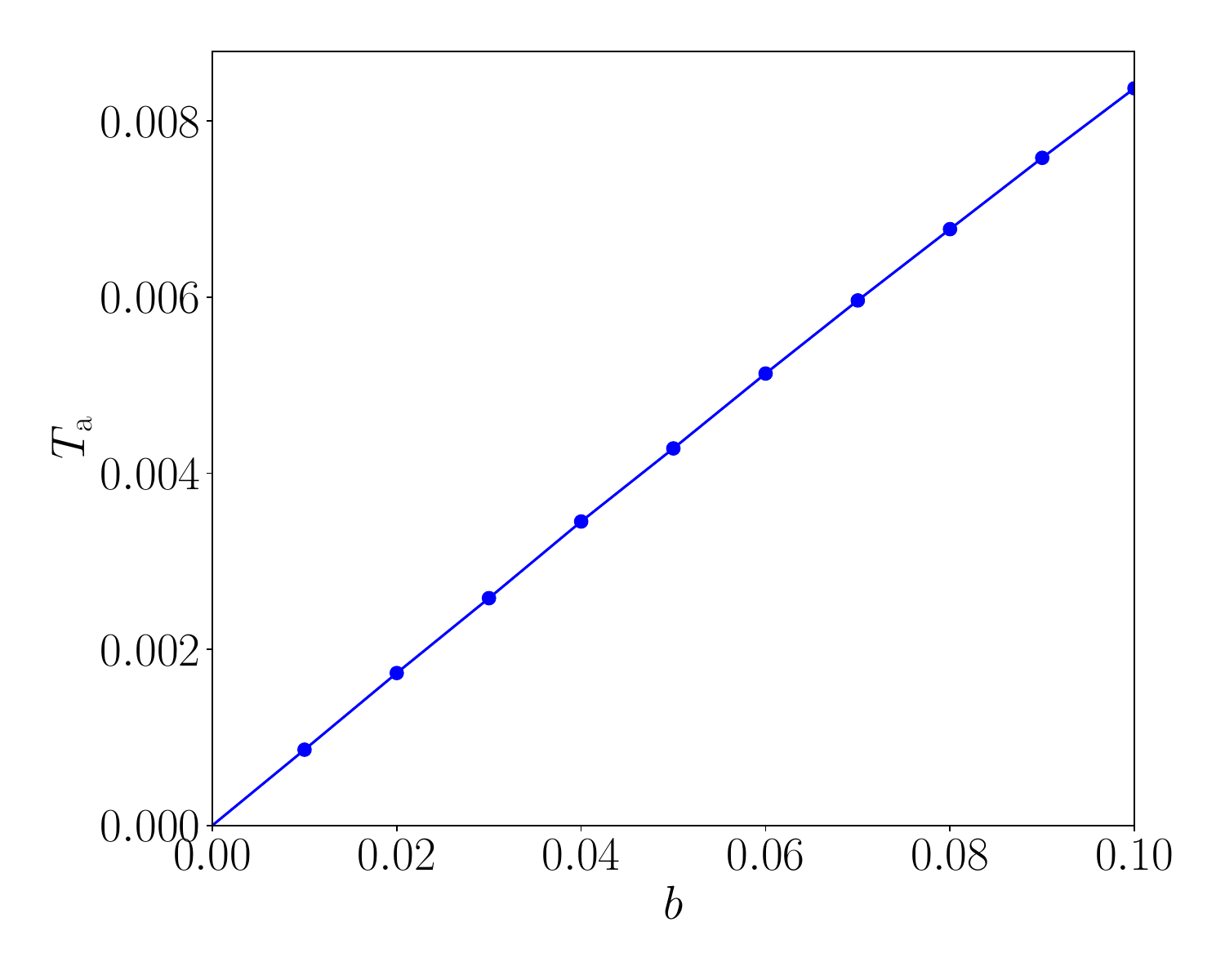}
  \caption{
    \label{fig:C_Td}
    \tm{
      The characteristic temperature $T_{\rm a}$ below which
      $(C(90.01^{\circ})-C(0^{\circ}))/C(0^{\circ}) \geq 2$,
      as a function of $b$.
      The solid line is a guide to the eye.
      }
  }
\end{figure}

\tm{    
    Figure~\ref{fig:C_Td} shows
    the characteristic temperature $T_{\rm a}$ below which
    $(C(90.01^{\circ})-C(0^{\circ}))/C(0^{\circ}) \geq 2$,
    as a function of $b$.
    We clearly see the linear dependence of $T_{\rm a}$
    with respect to the magnetic field $b$.
    We note that the difference arises at quite low temperatures.
    In fact for the case of $b=0.02$. $T_{\rm a} = 0.002$.
    So, if we take $J_{\gamma} = 100$ K for the Kitaev interaction,
    $T_{\rm a} = 0.2$ K.
    In order to observe the difference found in the experiment\cite{Tanaka2022}
    we need to measure the specific heat much lower temperatures.
    }

\section{Full Magnetic Field Effects on the Kitaev Model}
\label{sec:m_finite}
Now, we consider the full magnetic field effect on the Kitaev model.
In solving the self-consistent equations (\ref{sec:formalism:eq:sc_eq0})
and (\ref{sec:formalism:eq:sc_eq_gamma}),
we take the initial value of $m = \sqrt{m_x^2 + m_y^2 + m_z^2}$ from the HTE.
Specifically, we obtain it from the HTE of $m_{\gamma}$ ($\gamma= x, y, z$):
\be
   m_\gamma^{\rm HTE} = \frac{1}{4}\beta {b_\gamma} + O(\beta^3).
   \ee
   
\begin{figure}[htbp]
  \includegraphics[width=1 \linewidth, angle=0]{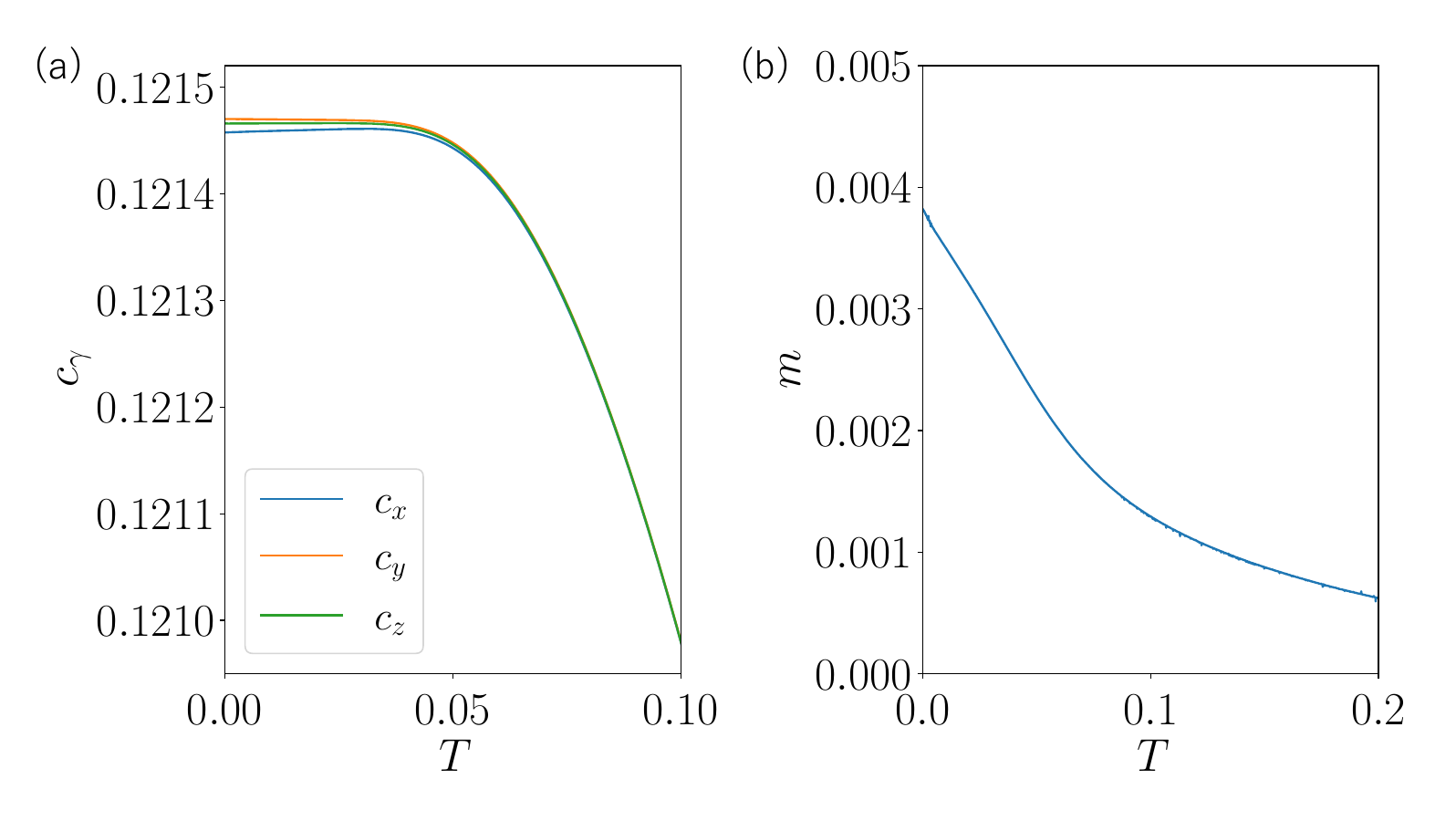}
  \caption{
    \label{fig:m:c_alpha_m_T}
    (a) Temperature dependence of the spin correlation functions $c_x$, $c_y$, and $c_z$. 
    (b) Temperature dependence of the magnetization $m$.
    Both (a) and (b) are computed for \tm{$b = 0.01$}
    and $\phi = 40^\circ$.
  }
\end{figure}

\begin{figure}[htbp]
  \includegraphics[width=0.9 \linewidth, angle=0]{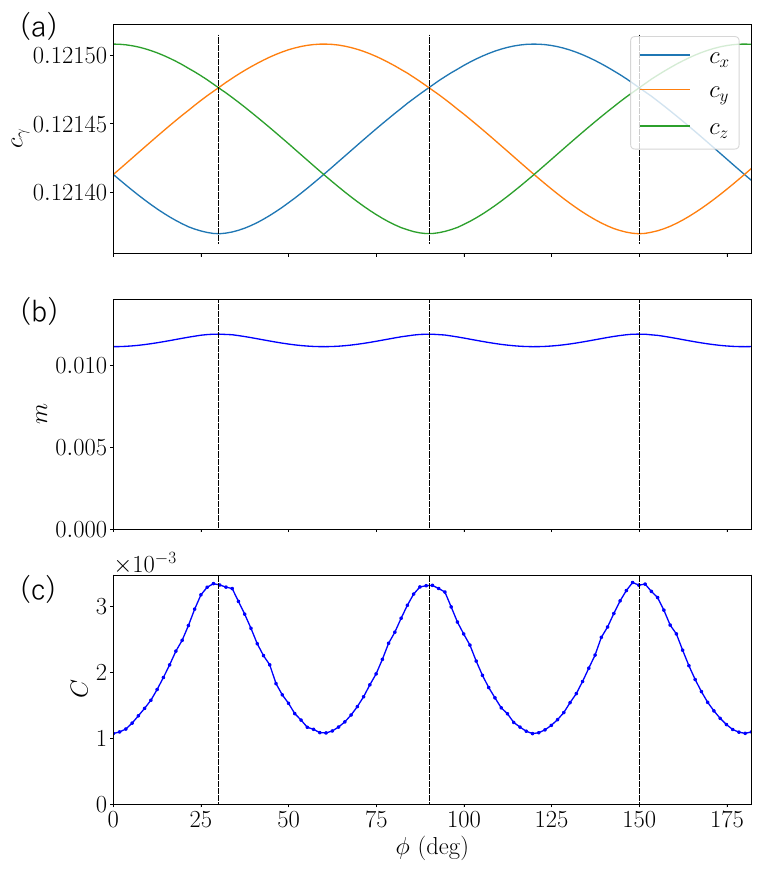}
  \caption{
    \label{fig:m:c_m_E_C_T0_05}
    \tm{
      (a) Spin correlation functions $c_x$, $c_y$, and $c_z$;
      (b) magnetization $m$;
      (c) specific heat, all plotted as functions of the angle $\phi$.
      In panel (c), the solid lines are guides to the eye.
      The vertical lines indicate specific angles as described earlier.
      The magnetic field is $b = 0.03$, and the temperature is $T = 0.001$.
      }
  }
\end{figure}

\begin{figure}[htbp]
  \includegraphics[width=0.9 \linewidth, angle=0]{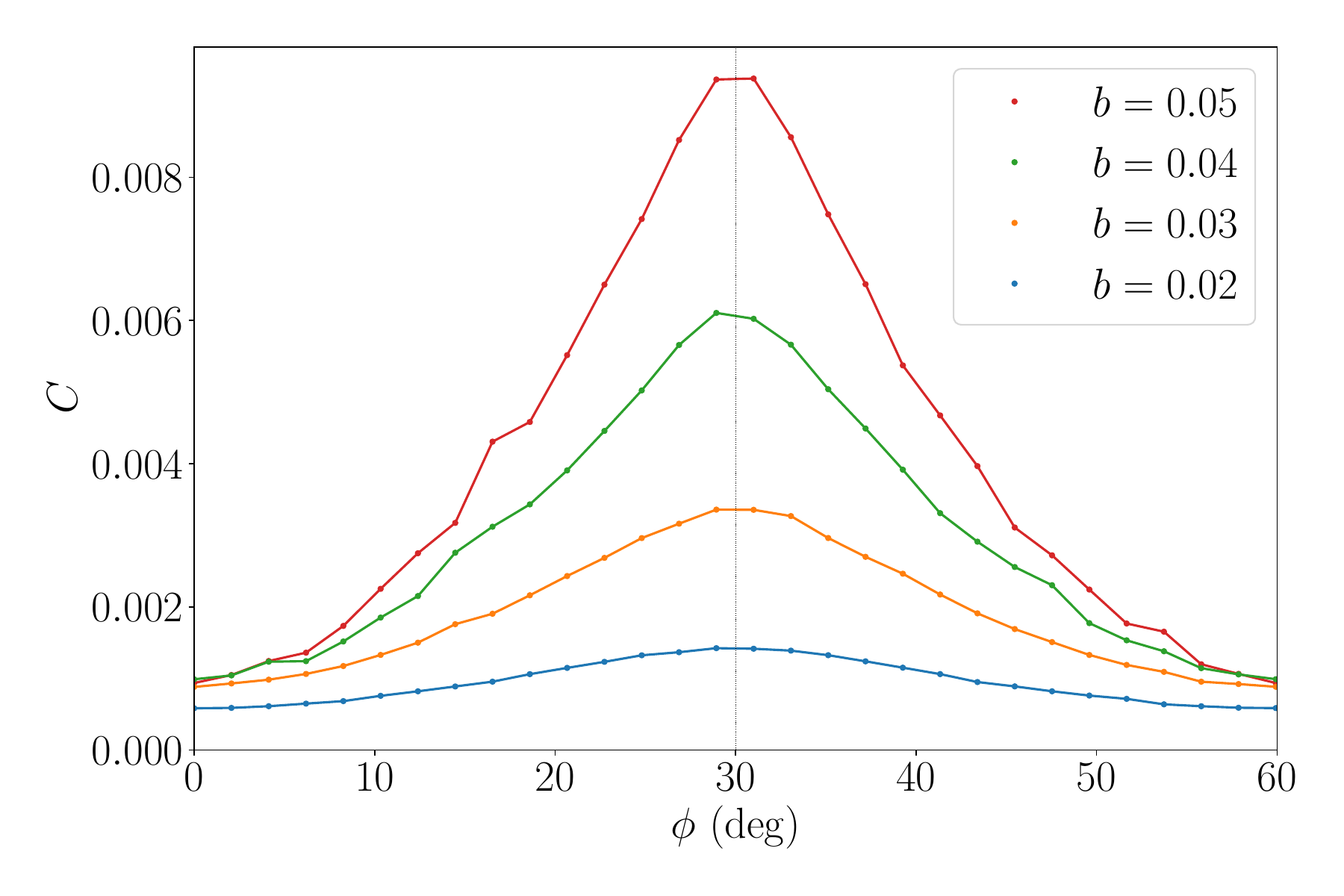}
  \caption{
    \label{fig:m:m_C_b}
    \tm{
      Angular dependence of the specific heat at $T = 0.01$
      for various magnetic field strengths.
      The vertical line indicates $\phi = 30^\circ$.
    }
  }
\end{figure}

\begin{figure}[htbp]
  \includegraphics[width=0.9 \linewidth, angle=0]{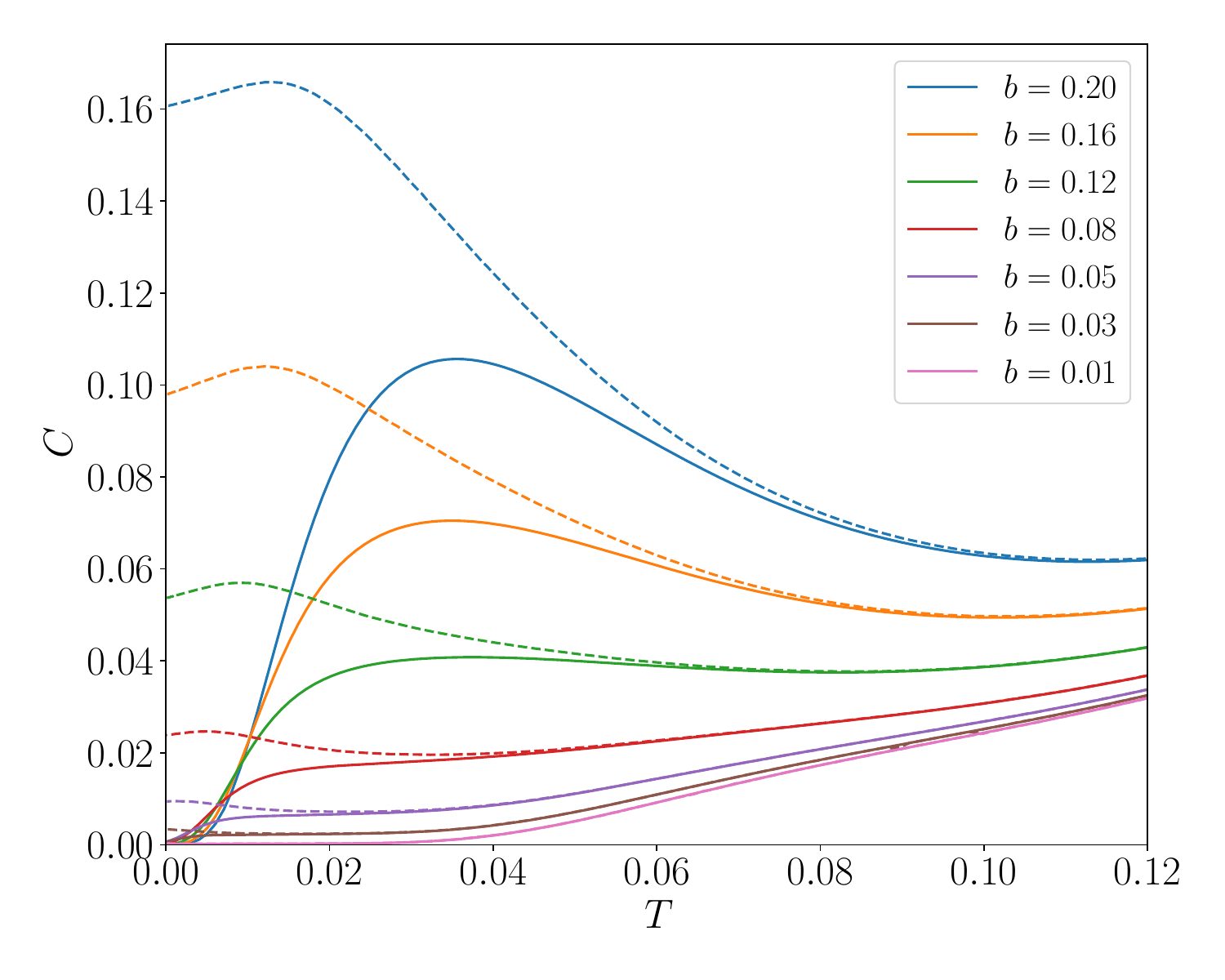}
  \caption{
    \label{fig:m:C_T_phi0_90}
    \tm{
      Temperature dependence of the specific heat $C$ for various magnetic field strengths
      applied along the $\phi = 0^\circ$ (solid lines)
      and $\phi = 90.01^\circ$ (dashed lines) directions.
      Note that $\phi = 90.01^\circ$ is used instead of exactly $\phi = 90^\circ$
      to avoid the spin inversion symmetry present
      in the $z$-component of the spins at $\phi = 90^\circ$.
      }
  }
\end{figure}

\begin{figure}[htbp]
  \includegraphics[width=0.9 \linewidth, angle=0]{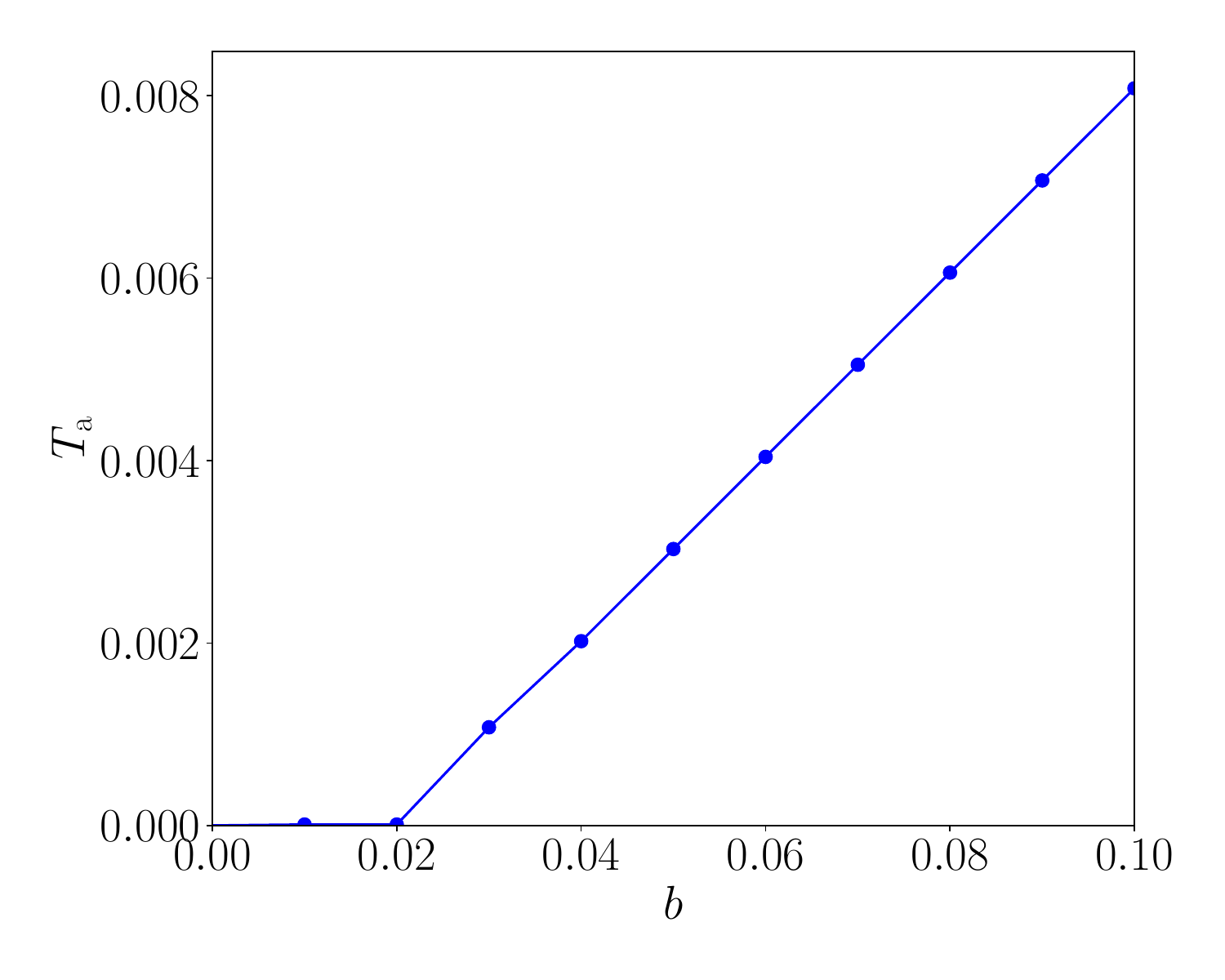}
  \caption{
    \label{fig:C_mag_Td}
    \tm{
      The characteristic temperature $T_{\rm a}$, defined as the temperature
      below which $(C(90.01^\circ) - C(0^\circ)) / C(0^\circ) \geq 2$,
      plotted as a function of magnetic field strength $b$.
      The solid line is a guide to the eye.      
      }
  }
\end{figure}

   Figure~\ref{fig:m:c_alpha_m_T} shows the temperature dependence of the spin correlation functions and the magnetization $m$.
   The temperature dependence of the correlation functions is similar
   to the case with $m = 0$.
   Figure~\ref{fig:m:c_m_E_C_T0_05} shows the angular dependence of
   the correlation functions,
   magnetization, and specific heat.
   The energy per site is defined as:
\be
E =  - \frac{1}{2} J \left( c_x + c_y + c_z \right) - m \left| {\bm{b}} \right|.
\ee
The specific heat is obtained by taking the temperature derivative of this energy.
The angular dependence of the correlation functions is similar to the $m = 0$ case,
where $c_\gamma$ attains its minimum values at the angles where ${\tilde b}_\gamma = 0$.
\tm{
  The magnetization, $m$, exhibts an oscillation with
  $60^\circ$ periodicity with a slight amplitude.
  The specific heat also exhibits a $60^\circ$ periodicity
  as in the case of $m=0$
  and
  the positions of the maxima and minima aligning
  with experimental observations \cite{Tanaka2022}.
  In Fig.~\ref{fig:m:m_C_b}, we show the angular dependence
  of the specific heat for $0 \le \phi \le 60^\circ$.
  The oscillation amplitude
  of the specific heat increases with increasing magnetic field strength.
  The qualitative behavior is similar to the $m = 0$ case:
  the minima continue to exhibit gap-like features,
  while the maxima do not show signatures of gapless Majorana fermions
  as seen in Fig.~\ref{fig:m:C_T_phi0_90}.
  Moreover, the angular dependence of the specific
  heat nearly vanishes for $b < 0.02$,
  since the characteristic temperature $T_{\rm a}$ becomes very small
  in this magnetic field range, as shown in Fig.~\ref{fig:C_mag_Td}.
}

\begin{figure}[htbp]
  \includegraphics[width=1 \linewidth, angle=0]{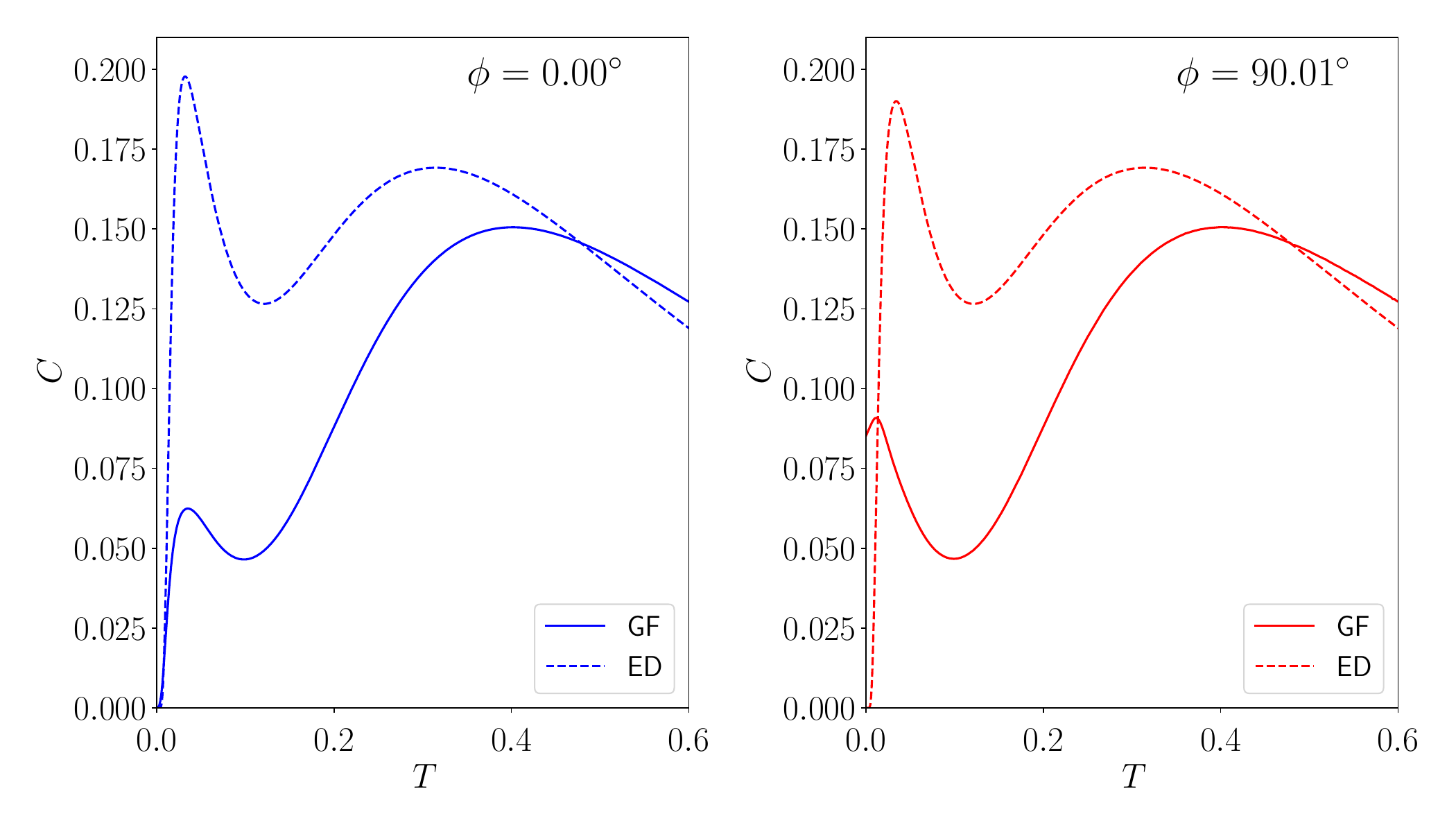}
  \caption{
    \label{fig:GF_ED}
    \tm{
      Temperature dependence of the specific heat at $\phi = 0^\circ$
      (left panel) and $\phi = 90.01^\circ$ (right panel) for $b = 0.15$,
      computed using the Green's function approach and exact diagonalization
      on a 13-site cluster.
      Both methods exhibit qualitatively similar behavior,
      showing a broad peak around $T \sim 0.3$–0.4 and an additional peak
      at lower temperatures.
      While the Green's function approach captures both features,
      it fails to reproduce the vanishing behavior at $\phi = 90.01^\circ$
      as $T \to 0$, highlighting its limitations at very low temperatures.
    }
  }
\end{figure}

\tm{
  We note that the specific heat at $\phi = 90.01^\circ$ does not approach zero as $T \rightarrow 0$.
  This is due to the fact that our Green's function approach becomes unreliable at low temperatures.
  To identify the temperature range where our method remains valid,
  we compare our results with those obtained from exact diagonalization
  on a cluster of three honeycombs (13 sites).
  The temperature dependence of the specific heat for $\phi = 0^\circ$
  and $\phi = 90.01^\circ$ at $b = 0.15$ is shown in Fig.~\ref{fig:GF_ED}.
  Although the system size used in the exact diagonalization
  is relatively small, the qualitative behavior is similar
  in both approaches.
  A broad peak appears around $T \sim 0.4$ in our calculation and
  around $T \sim 0.3$ in the exact diagonalization result.
  Both methods also show an additional peak at lower temperatures.
  While the Green's function approach captures this low-temperature peak
  at $\phi = 90.01^\circ$, consistent with the exact diagonalization result,
  it fails to approach zero as $T \rightarrow 0$.
  In this sense, our approach should be considered unreliable
  below the temperature at which the specific heat exhibits
  its lower-temperature peak.
  }

\section{Summary}
\label{sec:summary}
\tm{
  We have investigated the effect of a magnetic field on the Kitaev model
  using the equation of motion approach for the spin Green's function,
  considering both the case of suppressed magnetization ($m = 0$)
  and the case with finite magnetization ($m \neq 0$).
}

\tm{
  When magnetization is suppressed, the specific heat exhibits a clear $60^\circ$
  periodicity in its angular dependence, with the positions of the maxima and minima
  consistent with recent experimental
  observations in $\alpha$-RuCl$_3$ \cite{Tanaka2022}.
  A qualitative difference is observed in their temperature dependence:
  the minima exhibit gap-like behavior that may signal Majorana gap formation due to time-reversal symmetry breaking,
  while the maxima do not,
  deviating from the behavior expected for gapless Majorana fermions.
}

\tm{
  Furthermore, the characteristic onset temperature below which the angular
  dependence of the specific heat becomes visible increases linearly
  with the magnetic field.
  This implies the presence of a linear-in-field effect,
  independent of magnetization,
  which interferes with the isolation of Majorana fermion features.
}

\tm{
  When finite magnetization is included, the angular dependence of the specific heat remains, and the qualitative behavior is similar to the $m = 0$ case: the minima continue to exhibit gap-like features, while the maxima do not show signatures consistent with gapless Majorana fermions. Importantly, we find that the characteristic temperature below which this anisotropy becomes apparent increases linearly with the magnetic field strength and remains significantly lower than the temperature scale inferred from experiments. In particular, this onset temperature is nearly zero for $b \lesssim 0.02$, indicating that directional dependence is effectively unobservable in this regime. Additionally, the specific heat does not vanish as $T \to 0$, which reflects a known limitation of our Green's function approach at low temperatures. Comparison with exact diagonalization confirms that while our method captures qualitative trends at intermediate temperatures, it becomes unreliable near $T = 0$, especially in the presence of finite magnetization. Thus, our conclusions are restricted to the temperature regime above the low-temperature peak in the specific heat.
}

\tm{    
  Our analysis indicates that suppressing magnetization alone is insufficient
  to reveal quantum spin liquid behavior in the Kitaev model under a magnetic field.
  To explore the possible realization of the KSL,
  it is essential to consider the effects of additional interactions
  beyond the pure Kitaev model, such as antiferromagnetic Heisenberg
  or other non-Kitaev terms \cite{Chaloupka2010, Singh2012, Rau2014, Kim2016}.
  Alternatively, the antiferromagnetic Kitaev interaction itself may offer
  a promising route, as investigated
  in previous studies \cite{Zhu2018, Gohlke2018, Hickey2019}.
}

\tm{
  Extending our approach to include non-Kitaev interactions and applying it
  to the antiferromagnetic Kitaev model represent important directions
  for future research.
  Such efforts may further clarify the conditions under which Majorana fermion
  features become experimentally accessible.
}
\appendix
\section{Algorithm for Solving the Green's Functionn Equation}
\label{app:algorithm}
In this appendix, we describe how to solve the equations for the Green's functions in the form of Eq.~(\ref{eq:formalism:matrix_G_form}).  
In general, we need to solve the following equation for ${\bm G}(i\omega_n)$:
\be
i{\omega _n}{\bm{G}}(i\omega_n) = M{\bm{G}}(i\omega_n) + {\bm{c}},
\label{app:eq:G_eq}
\ee
where $M$ is an $m \times m$ matrix, ${\bm c}$ is an $m$-dimensional vector,
and ${\bm{G}}(i\omega_n)$ has the following form:
\be
   {\bm{G}}(i\omega_n) = \left( {\begin{array}{*{20}{c}}
{{G_1}(i\omega_n)}\\
{{G_2}(i\omega_n)}\\
 \vdots \\
{{G_m}(i\omega_n)}
\end{array}} \right).
   \ee
Let $\lambda_{\ell}$ and ${\bm v}_{\ell}$ ($\ell = 1,2,\dots,m$) be the eigenvalues and eigenvectors of the matrix $M$, respectively, satisfying:
\be
M{{\bm{v}}_\ell } = {\lambda _\ell }{{\bm{v}}_\ell }.
\ee
If the eigenvectors ${{\bm{v}}_\ell }$
are linearly independent, then ${\bm{G}}(i\omega_n)$ can be expanded as:
\be
   {\bm{G}}\left( {i{\omega _n}} \right)
   = \sum\limits_{\ell  = 1}^m {{C_\ell }\left( {i{\omega _n}} \right){{\bm{v}}_\ell }}.
   \label{app:eq:G_Cv}
   \ee
   The coefficient $C_\ell(i\omega_n)$ is determined by substituting this equation
   into Eq.~(\ref{app:eq:G_eq}).
   This yields:
   \be
{C_\ell }\left( {i{\omega _n}} \right) = \sum\limits_{\ell ' = 1}^m {\frac{{{{\left( {{V^{ - 1}}} \right)}_{\ell \ell '}}\left( {{\bm{v}}_{\ell '}^* \cdot {\bm{c}}} \right)}}{{i{\omega _n} - {\lambda _\ell }}}}.
\ee
Here, the components of the matrix $V$ are defined as:
${V_{\ell \ell '}} = {\bm{v}}_\ell ^* \cdot {{\bm{v}}_{\ell '}}$.
Thus, from Eq.~(\ref{app:eq:G_Cv}), we obtain:
\be
   {\bm{G}}\left( {i{\omega _n}} \right) = \sum\limits_{\ell  = 1}^m {\frac{{{{\bm{v}}_\ell }
         {{\left( {{V^{ - 1}}} \right)}_{\ell \ell '}}\left( {{\bm{v}}_{\ell '}^* \cdot {\bm{c}}} \right)}}{{i{\omega _n} - {\lambda _\ell }}}}.
   \ee

\begin{acknowledgments}
  We sincerely thank Dr. D. Kaib for his insightful comments,
  which helped us identify and correct an error in our previous calculation.
\end{acknowledgments}

\bibliography{../../../../refs/lib}
\end{document}